\documentclass[a4paper,prb,twocolumn,showpacs]{revtex4}
\usepackage{amsmath}
\usepackage{subfigure}
\usepackage{amsfonts}
\usepackage{ifthen}
\usepackage{ulem}
\usepackage[T1]{fontenc}
\usepackage{color}
\usepackage{pstricks}

\usepackage[dvips]{graphicx}
\usepackage[hypertex,breaklinks=true,colorlinks=false]{hyperref}
\usepackage[active]{srcltx} % For ``reverse search'': you click in the dvi file and you are redirected in the .tex source file

\newcommand{\pr}{Phys. Rev.}
\newcommand{\jpsj}{J. Phys. Soc. Jpn.}

\newcommand{\journaldoi}[5]{\href{http://dx.doi.org/#5}{#1\ {\bf #2}, #3 (#4)}}

\newcommand{\journal}[4]{
 \ifthenelse{\equal{#1}{pr}}{\journaldoi{\pr}{#2}{#3}{#4}{10.1103/PhysRev.#2.#3}}
{\ifthenelse{\equal{#1}{pra}}{\journaldoi{\pra}{#2}{#3}{#4}{10.1103/PhysRevA.#2.#3}}
{\ifthenelse{\equal{#1}{pre}}{\journaldoi{\pre}{#2}{#3}{#4}{10.1103/PhysRevE.#2.#3}}
{\ifthenelse{\equal{#1}{prd}}{\journaldoi{\prd}{#2}{#3}{#4}{10.1103/PhysRevD.#2.#3}}
{\ifthenelse{\equal{#1}{prl}}{\journaldoi{\prl}{#2}{#3}{#4}{10.1103/PhysRevLett.#2.#3}}
{\ifthenelse{\equal{#1}{prb}}{\journaldoi{\prb}{#2}{#3}{#4}{10.1103/PhysRevB.#2.#3}}
{\ifthenelse{\equal{#1}{rmp}}{\journaldoi{\rmp}{#2}{#3}{#4}{10.1103/RevModPhys.#2.#3}}
{\ifthenelse{\equal{#1}{arxiv}}{\href{http://arxiv.org/abs/#2.#3}{arXiv:#2.#3}}
{\ifthenelse{\equal{#1}{cond-mat}}{\href{http://arxiv.org/abs/cond-mat/#2}{cond-mat/#2}}
{\ifthenelse{\equal{#1}{jpsj}}{\journaldoi{\jpsj}{#2}{#3}{#4}{10.1143/JPSJ.#2.#3}}
{#1 {\bf #2}, #3 (#4)}}}}}}}}}}}
\begin{document}

\newcommand{\Z}{\mathbb{Z}}

\title{Thermal destruction of chiral order in a two-dimensional model of coupled trihedra}

\author{Laura Messio$^1$}\author{Jean-Christophe Domenge$^2$}\author{Claire Lhuillier$^1$}
\author{Laurent Pierre$^1$}
\author{Pascal Viot$^1$}\author{Gr\'egoire Misguich$^3$}
\affiliation{$^1$Laboratoire de Physique Th\'eorique de la Matière condens\'ee,
UMR 7600 C.N.R.S., Universit\'e Pierre-et-Marie-Curie, Paris VI, France. }

\affiliation{$^2$Department of Physics and Astronomy and \mbox{Center for
Condensed Matter Theory},\ Rutgers University,\ Piscataway,\ NJ
08854-8019}

\affiliation{$^3$Institut de Physique Théorique,
CEA, IPhT, F-91191 Gif-sur-Yvette, France.
CNRS, URA 2306.}

\begin{abstract}
We introduce  a minimal   model  describing the physics   of classical
two-dimensional (2D) frustrated Heisenberg  systems, where spins order
in  a non-planar way   at $T=0$.   This  model, consisting  of coupled
trihedra (or Ising-$\mathbb{R}P^3$  model), encompasses Ising (chiral)
degrees   of    freedom,     spin-wave  excitations    and      $\Z_2$
vortices. Extensive Monte Carlo simulations show that the $T=0$ chiral
order disappears  at  finite   temperature  in  a   continuous   phase
transition in the   $2D$ Ising universality class,  despite misleading
intermediate-size effects observed at  the transition. The analysis of
configurations reveals  that short-range spin  fluctuations and $\Z_2$
vortices  proliferate near the  chiral   domain walls explaining   the
strong renormalization of the   transition temperature. Chiral  domain
walls  can themselves carry an  unlocalized $\Z_2$ topological charge,
and vortices  are   then  preferentially paired  with charged   walls.
Further, we conjecture  that  the anomalous size-effects  suggest  the
proximity of the   present model to a  tricritical  point.  A body  of
results is  presented, that  all  support this claim:  (i) First-order
transitions   obtained by Monte   Carlo simulations on several related
models (ii)   Approximate  mapping  between  the Ising-$\mathbb{R}P^3$
model and  a dilute Ising model (exhibiting  a tricritical point) and,
finally,   (iii)  Mean-field  results   obtained  for  Ising-multispin
Hamiltonians, derived   from  the high-temperature expansion   for the
vector spins of the Ising-$\mathbb{R}P^3$ model.
\end{abstract}

\date{\today}

%05.50.+q %05.50.+q Lattice theory and statistics (Ising, Potts, etc.)
%75.10.Hk %75.10.Hk Classical spin models
%75.30.Kz %75.30.Kz Magnetic phase boundaries (including magnetic transitions, metamagnetism, etc.)


\maketitle

%\tableofcontents

\section{Introduction}
On bipartite lattices, the energy of the classical Heisenberg, as well
as XY, antiferromagnet is minimized  by collinear spin configurations.
Any two  such ground states can be  continuously transformed  into one
another by a  global spin rotation.  By  contrast, it is  quite common
that the ground state  manifold of {\it frustrated}  magnets comprises
several connected components,  with respect to global spin  rotations,
that    transform  into one  another   under  discrete  symmetry only.
Examples include Villain's fully frustrated XY model,~\cite{villain77}
the      $J_1-J_2$      Heisenberg     model     on     the     square
lattice,~\cite{ccl90,weber03}   the  $J-K_4$ model  on the  triangular
lattice,~\cite{momoi97}  the   $J_1-J_3$     model   on   the   square
lattice,~\cite{cs04}  and the    $J_1-J_2$   model  on the    kagom\'e
lattice.~\cite{domenge05,domenge08}

The Mermin-Wagner   theorem~\cite{mermin66} forbids  the   spontaneous
breakdown of   continuous symmetries, such as  spin  rotations, at any
$T>0$  in  two  dimensions (2D).    However, as was   first noticed by
Villain,~\cite{villain77}  the   breakdown of the  discrete symmetries
relating  the different   connected   components of the ground   state
manifold   may indeed    give   rise  to  finite   temperature   phase
transition(s).  Such transitions have  been evidenced numerically in a
number of frustrated  systems, with either XY~\cite{numerics_FFXY}  or
Heisenberg spins.~\cite{momoi97,weber03,domenge05,domenge08}

We are interested   in a particular  class of   models with Heisenberg
spins,  where   the ground state     has {\it  non-planar}  long-range
order.~\cite{momoi97,domenge05,domenge08} In   this    case the ground
state is labeled by an $O(3)$ matrix. Hence the ground state manifold
is   $O(3)=SO(3)\times\Z_2$  which breaks  down   into  two copies  of
$SO(3)$.  The two connected components, say  1 and 2, are exchanged by
a global  spin inversion ($\vec S_i\to-\vec  S_i$) and may be labeled
by  opposite scalar chiralities $\vec  S_i \cdot (\vec S_j \times \vec
S_k)$.  Hence we introduce  a local Ising variable $\sigma(\vec r)=\pm
1$ which measures whether the spins around $\vec r$ have the chirality
of sector  1  or 2.  At  $T=0$ the  chiralities  $\sigma(\vec  r)$ are
long-range ordered and the ground state belongs  to a given sector. On
the other hand,   at  high  enough temperature  the   system is  fully
disordered.  Hence, on very  general grounds we expect the spontaneous
breakdown  of    the  spin    inversion   symmetry,  associated     to
$\langle\sigma(\vec r)\rangle\neq0$, at some intermediate temperature.

Further,    from  the  standpoint     of  Landau-Ginzburg  theory, one
anticipates a critical transition in  the 2D Ising universality class.
However,   of     the     two   relevant      models     studied    so
far,~\cite{momoi97,domenge05,domenge08} none shows the signature of an
Ising  transition. Instead,  as   was pointed out   by  some of  us in
Ref.~\onlinecite{domenge08},  the existence of underlying (continuous)
spin  degrees of freedom  complicates   the naive Ising scenario,  and
actually drives the chiral transition towards first-order.

To   get a better    sense of this    interplay  between discrete  and
continuous degrees of freedom,  it is useful  to remember that  in 2D,
although  spin-waves disorder  the spins  at  any $T>0$, the spin-spin
correlation    length   may be  huge     ($\xi\sim\exp(-A/T)$) at  low
temperature,~\cite{bz76}  especially in frustrated  systems. Hence, it
is likely that  the effective, spin-wave mediated, interaction between
the emergent Ising degrees of freedom extends significantly beyond one
lattice spacing, even at finite temperature and in 2D.

Further, we  point  out that the excitations  built  on the continuous
degrees of  freedom are not necessarily  limited to spin-waves.  To be
more    specific,  if the  connected components    of the ground state
manifold are  not {\it simply} connected,  as is the case for $SO(3)$,
then   there  also exists   defects   in  the  spin  textures.   Here,
$\Pi_1(SO(3))=\Z_2$ implies that $\Z_2$ point defects (vortices in 2D)
are topologically stable.   Clearly these  additional excitations  may
also affect the  nature of   the  transition associated  to the  Ising
degrees of    freedom.     In   fact,  it   was      shown   on    one
example~\cite{domenge08}  that the  first-order  chiral transition  is
triggered by the proliferation of these defects.

In this paper we aim at clarifying the nature of the interplay between
the different  types of excitations  found in  Heisenberg systems with
non-planar  long-range order at $T=0$.  Note  that the associated unit
cell is typically quite large,  which severely limits the sample sizes
amenable to simulations. Hence, we introduce a  minimal model with the
same physical   content  as   the    frustrated models  studied     in
Refs~\onlinecite{momoi97,domenge05,domenge08}.

As was already mentioned, in the above frustrated spin systems, the spin
configuration at $T=0$ is entirely described by an $O(3)$ matrix, or
equivalently a trihedron in spin space. At low temperature, the spin
long-range order is wiped out by long wavelength spin waves, but from
the considerations above we anticipate that, at low enough temperatures,
the description in terms of trihedra in spin space still makes sense, at
least locally. To be more specific, we assign three unit vectors
$\vec{S_i}^a$ ($a=1,2,3$) to every site  $i$ of the square lattice,
subject the orthogonality constraint
\begin{equation}
\vec{S_i}^a \cdot\vec{S_i}^b=\delta^{a,b},
\label{eq:ortho}
\end{equation}
and we assume the following interaction energy
\begin{equation}
E=-\sum_{a=1}^3 \sum_{\langle i,j\rangle} \vec{S_i}^a \cdot \vec{S_j}^a.
\label{eq:energy}
\end{equation}
Hence we  consider  three  ferromagnetic Heisenberg  models,   tightly
coupled through the rigid  constraint~(\ref{eq:ortho}). At $T=0$,  the
energy is minimized by   any configuration with all trihedra  aligned,
and the manifold of  ground states is $O(3)$,  as desired. This  alone
ensures the  existence of the three  types  of excitations: i) $SO(3)$
spin waves   (corresponding  to the   rotation of  the  trihedra), ii)
$SO(3)$ vortices,  and  iii)    Ising (chiral) degrees    of  freedom,
corresponding   to   the   right     or  left-handedness      of   the
trihedra.\footnote{This model does not  encompass the breathing  modes
of the trihedra that are present  in the full frustrated spin systems:
we suspect that these modes would not change qualitatively the present
picture.}

The present study is devoted to the  model  defined by
Eqs.~\ref{eq:ortho} and~\ref{eq:energy}.

In Section~\ref{sec:basics} we reformulate this model more conveniently
in terms of chiralities   and four-dimensional (4D) vectors, yielding
the so-called Ising-$\mathbb{R}P^3$ model.  For clarity we first
consider a simplified version of this model where the chirality
variables are frozen, and we use this setup to detail our method to
detect vortex cores and analyze their spatial distribution.

In section~\ref{sec:num}, we return to the full Ising-$\mathbb{R}P^3$
model and evidence the order-disorder transition of the Ising variables
at finite-temperature. The nature of the transition is asserted by a
thorough finite-size analysis using Monte-Carlo simulations. To clarify
the nature of the interplay between discrete and continuous degrees of
freedom we perform a microscopic analysis of typical
configurations.

For the most part, the remainder of our work originates from the
observation of peculiar intermediate size effects at the transition.
This leads us to argue that the present model lies close to a
tricritical point in some parameter space.

To support our claim we first introduce and  perform Monte Carlo simulations
on two modified versions of our model that i) preserve the $O(3)$ manifold
of ground states, and ii) lead to a first-order transition of the Ising
variables.

This is  further  elaborated on in  section~\ref{sec:potts},  where we
draw an analogy  between   the  Ising-$\mathbb{R}P^3$ model   and  the
large~$q$ Potts  model.  Another analogy, this time  to a dilute Ising
model, is drawn in section~\ref{sec:diluted},  where we argue that the
regions   of strong misalignment of   the  trihedra, near Ising domain
walls,  can be treated as ``depletions''  in the texture formed by the
4D-vectors.

Finally, in Section~\ref{sec:HT} we take another route and trace out the
continuous degrees of freedom perturbatively, resulting in an effective
model for the Ising variables, that we proceed to study at the
mean-field level.

\section{Basics of the model} \label{sec:basics}
\subsection{Ising-$\mathbb{R}P^3$ formulation}
The model defined by Eqs.~\ref{eq:ortho} and \ref{eq:energy} can be
conveniently reformulated as an Ising model coupled to a four-component
spin system with biquadratic interactions. Indeed, every trihedron is
represented by an $SO(3)$ matrix $M_i$ and a chirality $\sigma_i=\pm1$:
\begin{equation}
\vec{S_i}^{1}  =  \sigma_i   M_i \left[\begin{array}{c}1   \\  0  \\
  0\end{array}\right]    ,\,      \vec{S_i}^{2}   =     \sigma_i  M_i
  \left[\begin{array}{c}0 \\  1 \\ 0\end{array}\right] ,\, \vec{S_i}^{3}
    = \sigma_i M_i \left[\begin{array}{c}0 \\ 0 \\ 1\end{array}\right].
\label{eq:SchiM}
\end{equation}
Using these variables, Eq.~\ref{eq:energy} reads \
\begin{equation}
E=-\sum_{\langle i,j\rangle}  \sigma_i\sigma_j{\rm Tr}\left[ (M_i)^t
M_j\right]
\label{eq:energy_trace}
\end{equation}
The isomorphism between  $SO(3)$ and $SU(2)/\left\{1,-1\right\}$, maps a
rotation $M(\theta,\vec n)$ of  angle $\theta$ about the  $\vec n$ axis
onto the pair of  $SU(2)$ matrices $\pm \exp (i\frac{\theta}{2}\vec n
\cdot \vec \sigma)$, where  the components of $\vec \sigma$ are the
Pauli matrices. The  latter can be written using two opposite 4D real
vectors $\pm \vec v=\pm(v_0,v_x,v_y,v_z)$ with $\vec v^2=1$:
\begin{eqnarray}
\exp \left(i\frac{\theta}{2} \vec n \cdot \vec \sigma\right) =
\left[\begin{array}{cc} v_0+iv_z & iv_x+v_y \\ iv_x-v_y & v_0-iv_z
\end{array} \right] \\ v_0 =\cos(\theta/2) \,\,,\,\, v_a =n_a
\sin(\theta/2).
\end{eqnarray}

Irrespective of the local (arbitrary) choice of representation $\pm
\vec v_i$,  one has
\begin{equation}
{\rm Tr}\left[ (M_i)^t  M_j\right]=4(\vec v_i \cdot \vec v_j)^2-1
\end{equation}
whence the energy reads
\begin{equation}
E = -\sum_{\langle i,j\rangle} \sigma_i\sigma_j\left(4(\vec v_i \cdot
\vec v_j)^2-1 \right).
\label{eq:energy_vv2}
\end{equation}

\subsection{$\Z_2$ vortices in the fixed-chirality limit}
\label{sec:fixed-chir-limit}
Here we first consider the simplified case where the Ising degrees of
freedom  are  frozen to, say, $\sigma_i=+1$. In this limit we recover
the so-called  $\mathbb{R}P^3$ model,~\footnote{$\mathbb{R}P^n$ is   the
real projective  space.  It    is   formed by  taking the quotient  of
$\mathbb{R}^{n+1}-\{0\}$ under the  relation of equivalence $x\sim
\lambda x$  for  all  real numbers $\lambda\neq0$, or equivalently,   by
identifying    antipodal points   of    the unit    sphere,  $S^n$, in
$\mathbf{R}^{n+1}$.  $\mathbb{R}P^2$ was introduced  in the context of
liquid          crystals~\cite{RP2_old}            (see           also
Refs.~\onlinecite{lammert93-95,caracciolo93}). ~\cite{lammert93-95,caracciolo93}
}  which contains both spin-waves  and  $\Z_2$ vortices, and describes
interacting $SO(3)$ matrices.  Together  with related frustrated  spin
models (Heisenberg model on the  triangular lattice for instance), the
$\mathbb{R}P^3$   model has been the   central subject of  a number of
studies focusing  on  a putative binding-unbinding   transition of the
$\Z_2$                 vortices             at                  finite
temperature.~\cite{km84,ks87,dr89,adj90,kz92,adjm92,sy93,sx95,caracciolo93,hasenbusch96,wae00,cadm01}
This is a delicate and controversial issue, which  is not essential to
the present  work.  Instead, we  merely discuss some properties of the
vortex configurations that will be useful for comparison with the full
$O(3)$ (or Ising-$\mathbb{R}P^3$) model.
\begin{figure}
\subfigure[]{\includegraphics[width=2cm]{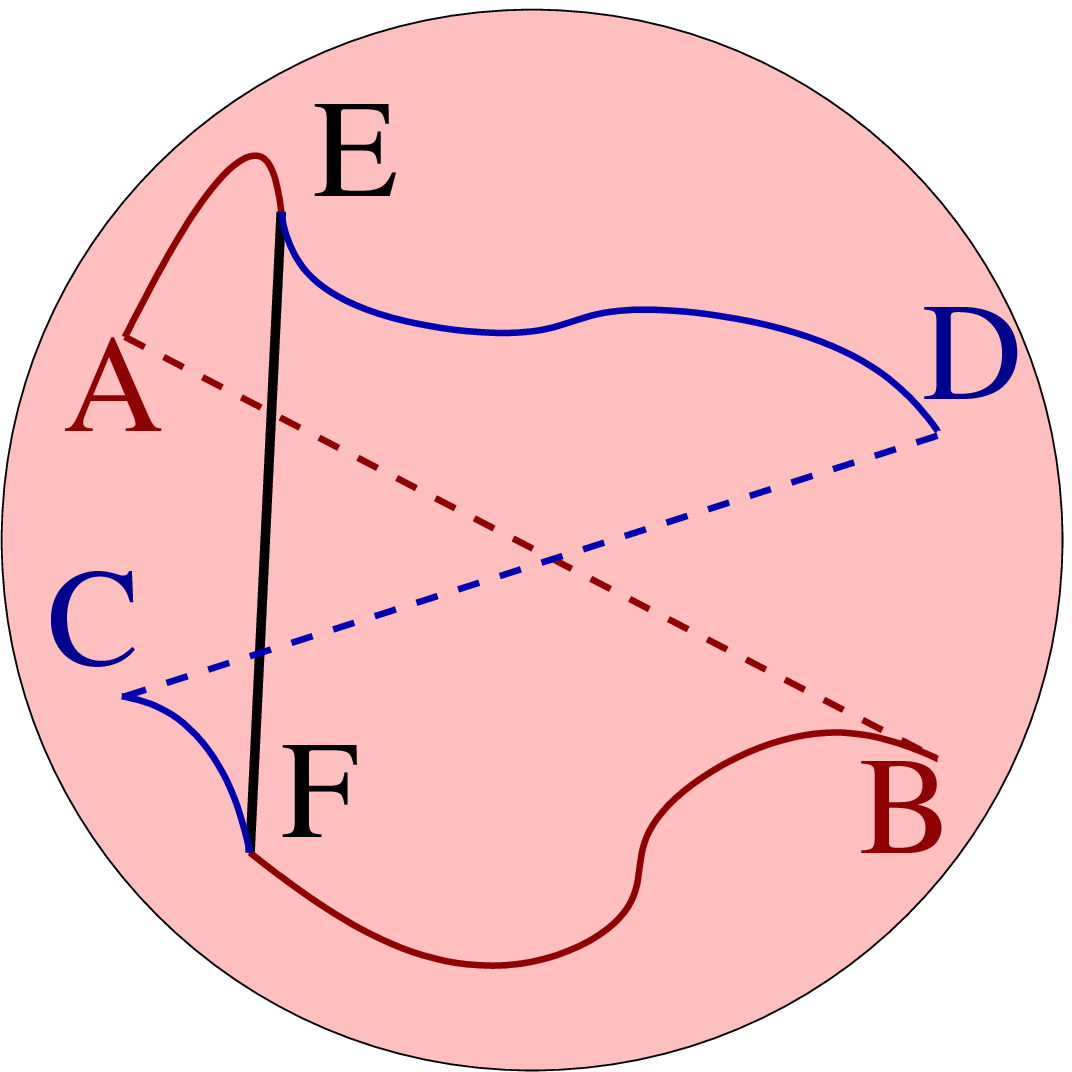}}
\subfigure[]{\includegraphics[width=2cm]{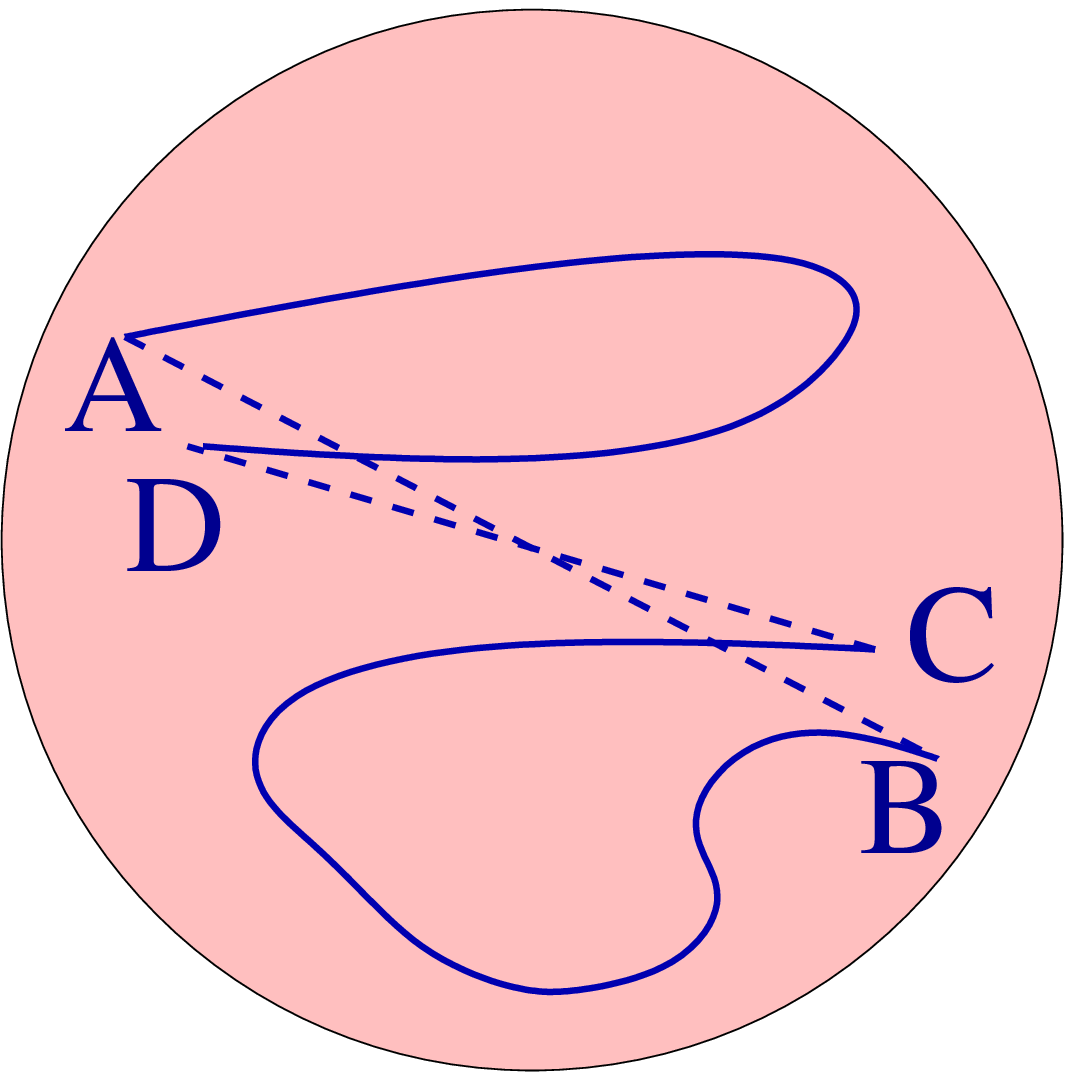}}
\subfigure[]{\includegraphics[width=2cm]{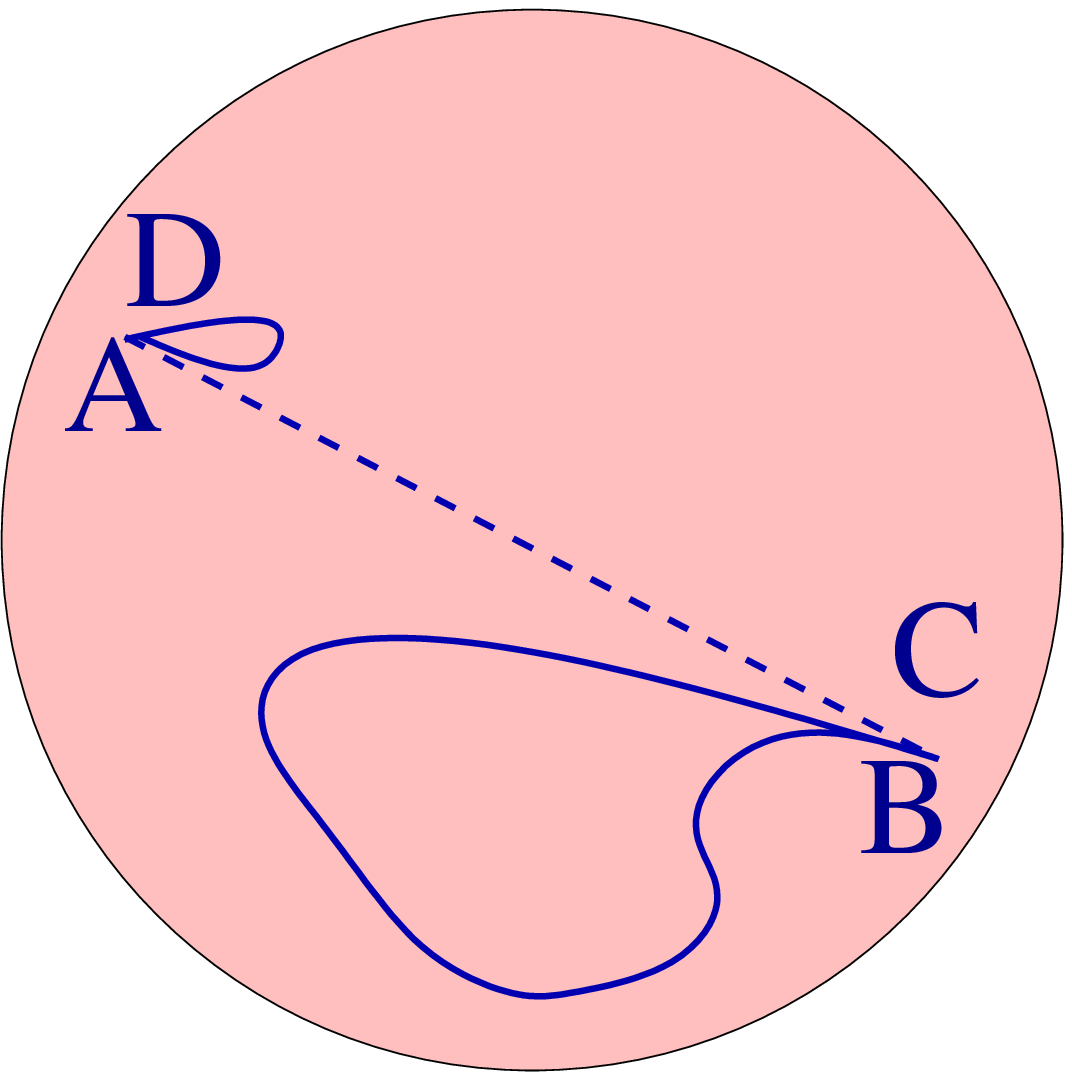}}
\subfigure[]{\includegraphics[width=2cm]{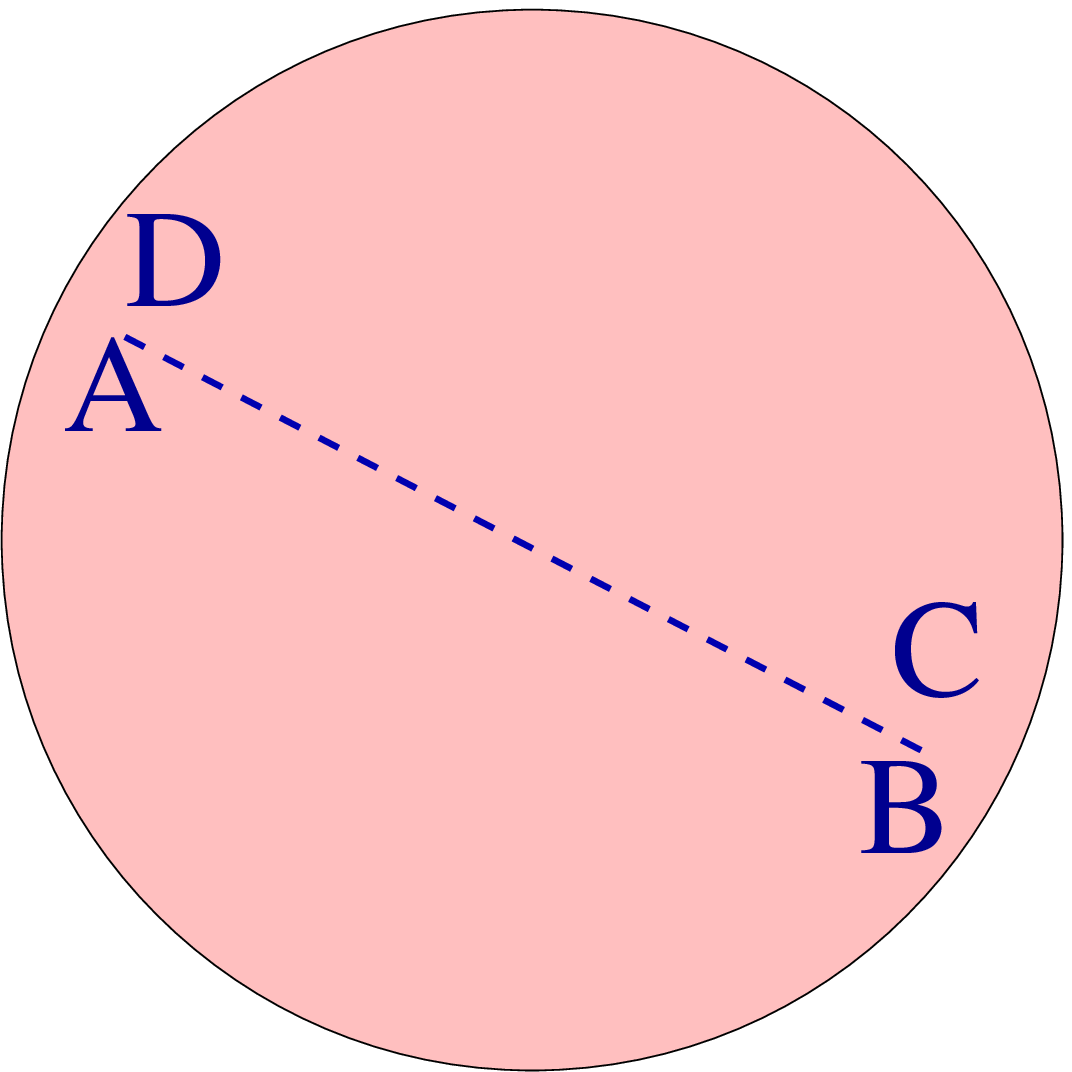}}
\caption{(Color online).  $SO(3)$ can be represented by a ball of radius
$\pi$.  A rotation of angle $\theta \in [0,\pi]$ around  the $\vec{n}$
axis is represented in  this ball by    the extremity of  the vector
$\theta \vec{n}$.  Two   opposite points of the   surface (joined by a
dashed line) represent the same rotation.  A  loop with an odd number of
crossings, such as $ABFEA$ or $CDEFC$, cannot be continuously shrunk to
a point. This is not the case when the number of crossings is even: a)
$ABCDEA$ crosses  the surface both at (AB) and (CD). b)-c) It can be
continuously deformed to collapse points $B$ on $C$, and $A$
on $D$.  d) $AD$ and $BC$ become contractible closed loops, that
collapse on identical points.}
\label{fig:vortex}
\end{figure}

To locate the topological point defects (vortex cores) we resort to the
usual procedure:   Consider a closed   path,  $\mathcal{L}$, on the
lattice, running         through  sites ${i_0},\cdots,{i_{n-1}},{i_n
\equiv i_0}$.  $\mathcal{L}$ induces a loop $\mathcal{C_L}$ in  the
order parameter space,   defined       by the     matrices
$M_{i_0},\cdots,M_{i_{n-1}},M_{i_n}$ (in this space the   path between
$M_{i_k}$ and $M_{i_{k+1}}$ is defined as the ``shortest one'').  The
homotopy class of $\mathcal{C_L}$ is an element of $\Pi_1$. If it is the
identity, then the $\mathcal{C_L}$ loop  is contractible.  Otherwise,
$\mathcal{L}$ surrounds (at least) one topological defect.~\cite{mermin}

In  the    $\mathbb{R}P^3$    model,   the  fundamental      group  is
$\Pi_1(SO(3))=\Z_2$, so that the topological charge can take only two
values: the homotopy class of  $\mathcal{C_L}$ correspond to the parity
of   the     number     of   point   defects enclosed in $\mathcal{L}$.
This is to be contrasted with the better-known $SO(2)$ vortices,
associated, for instance, to the $XY$ model: the latter carry an {\it
integer} charge ($\Pi_1(SO(2))=\Z$) while the former carry merely a {\it
sign}.

The number of $\Z_2$ vortices of a given configuration is obtained by
looking   for  vortex cores    on  each elementary plaquette of the
lattice. The vorticity $\Omega(p)$  of a square plaquette  $p$ is
computed  using  by    mapping $SO(3)$    to $\mathbb{R}P^3=
S^3/\mathbb{Z}_2$: on every site $i$ we arbitrarily  choose one of the
two equivalent representations $\pm\vec v_i$ of the local $SO(3)$ matrix
and compute
\begin{equation}
\Omega(p)=\prod _{\square_p} \text{sign}(\vec v_i.  \vec v_j),
\label{eq:1}
\end{equation}
where $i$  and $j$  are  nearest neighbors  and the product runs over
the four edges of ${\square_p}$.
The associated closed loop in $SO(3)$ is non-contractible when the
plaquette hosts a vortex core, and is identified by $\Omega(p)=-1$. Note
that $\Omega(p)$ is a ``gauge invariant'' quantity, {\it i.e.} it is
{\it independent} of the local choice of representation $\pm\vec v_i$,
as it should be.

\begin{figure}
\includegraphics[angle=-90,width=8cm]{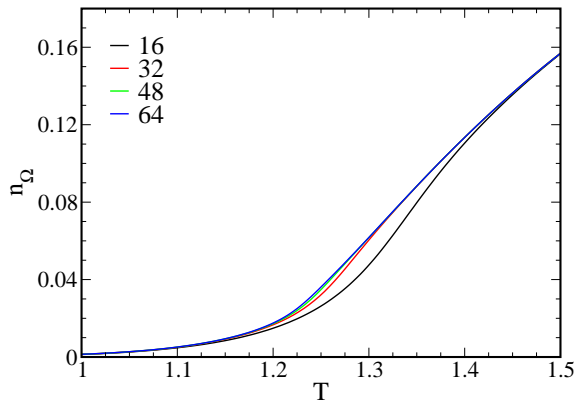}\\
\includegraphics[width=8cm]{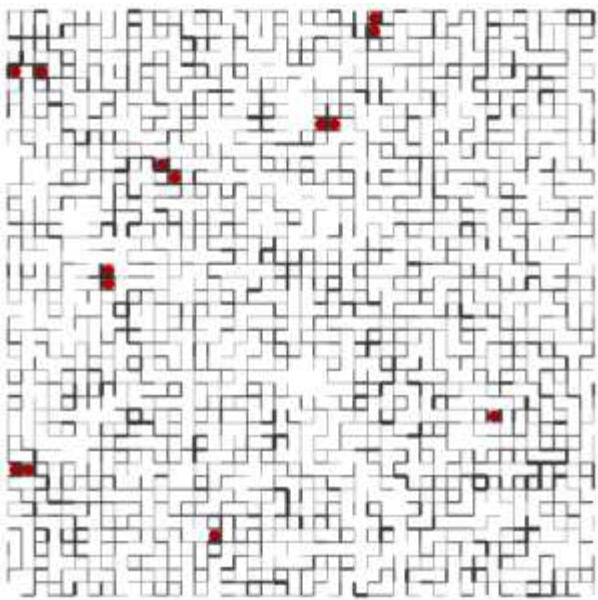}
\caption{(Color online) Uniform frozen chirality model (see text
paragraph  \ref{sec:fixed-chir-limit}).  Top:  Vortex density $n_\Omega$
versus    temperature     $T$, from $L=16$ to $L=64$, from   bottom
to   top.  Bottom: Thermalized  configuration  at
$T\simeq 1.1$ on a system  of  size $L=72$ (only  a  45$\times$45 square
snippet  is shown).  Square plaquettes carrying a  $\Z_2$ vortex core
are  denoted by a (red) bullet.  The width of  each bond $(i,j)$ is
proportional to $B_{ij}=1-(\vec v_i\cdot \vec v_j)^2$ (by slices of
$1/8$). Bonds with $B_{ij}\leq 1/8$ are not drawn.}
\label{fig:chi1}
\end{figure}

We performed a Monte Carlo simulation of the $\mathbb{R}P^3$ model using
a Wang-Landau algorithm, detailed in
Appendix~\ref{sec:monte-carlo-algor}.  In the upper panel of
Fig.~\ref{fig:chi1} we plot the  vortex      density $n_\Omega$, defined
as the number of plaquettes hosting a vortex core  divided by the total
number     of plaquettes.
Vortices are seen to appear and the  density increases upon increasing
the    temperature from   $T\simeq1$.  However,  no  critical behavior
(scaling) is  observed upon increasing the  system size. The latter is
also true of other simple thermodynamic quantities, such as the energy
or the specific heat, although the latter is maximum when the increase
in   $n_\Omega$ is  steepest,   at $T\simeq1.3$.~\footnote{Overall our
simulations show  no    obvious   signature of  a    vortex  unbinding
transition. However, more conclusive arguments require the computation
of   more     vortex-sensitive    quantities, such      as  the   spin
stiffness~\cite{cadm01},    or       area  versus   perimeter  scaling
laws,~\cite{km84} which are numerically very demanding.}

A   typical   configuration   is   shown   in   the   lower  panel  of
Fig.~\ref{fig:chi1}, at a temperature $T=1.1$, about $20\%$ lower than
that of the maximum of the specific heat.  Plaquettes hosting a vortex
core are  indicated by a (red)  bullet. Note that $n_\Omega$ is rather
small at  this temperature,  and  that all vortices  are {\it paired},
except for two, evidencing the strong binding of the vortices.

In the same figure the width of every bond $(i,j)$ is proportional to
$B_{ij}=1-(\vec{v}_i\cdot\vec{v}_j)^2$ and indicates the relative
orientation of the two 4D-vectors $\vec v_i$ and $\vec v_j$.  $B_{ij}=0$
(no segment) corresponds to \textit{parallel} 4D-vectors (or identical
trihedra), which minimizes the bond energy ( $E_{ij}=-3$). In
particular, all bonds have $B_{ij}=0$ at $T=0$. On the contrary,
$B_{ij}=1$ (thick black segment) indicates a maximally frustrated bond
($E_{ij}=+1$) with \textit{orthogonal} 4D-vectors (the two trihedra
differ by a rotation of angle $\pi$).  Figure~\ref{fig:chi1} shows that
the vortex cores are located in regions of enhanced short-range
fluctuations of the continuous variables (represented by thick bonds),
but that the converse is not necessarily true.

\section{Numerical simulations of the Ising--$\mathbb{R}P^3$ model}
\label{sec:num}
We now return to the full Ising-$\mathbb{R}P^3$ model, defined in
Eq.~\ref{eq:energy_vv2}.  We report results of Monte Carlo  simulations,
using the Wang-Landau algorithm described in Appendix
\ref{sec:monte-carlo-algor}, for  linear sizes up to $L=88$ with
periodic boundary   conditions.
\subsection{Specific heat and energy distribution}
\label{sec:energy_distrib}
\begin{figure}
\includegraphics[angle=-90,width=7cm]{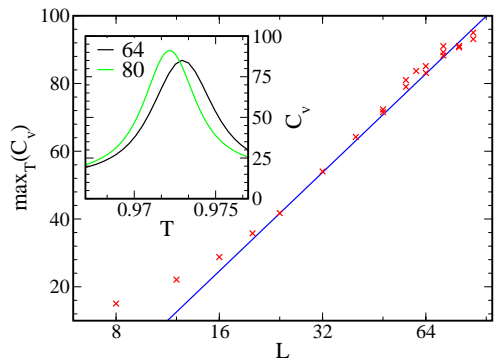}
\caption{(Color online)  Maximum value of the specific heat as a
function of system size $L$. The scaling at large $L$ is correctly
reproduced by the  affine log fit $A+B\ln(L)$ (straight line).  Inset:
specific heat $C_v$ versus  temperature $T$ for $L=64$ and $80$ from
right to left.}
\label{fig:cv}
\end{figure}

Once the Ising degrees of freedom are relaxed, the maximum of the
specific heat diverges with the system size, indicating a phase
transition. Further, the scaling as $\sim\log(L)$ (Fig.~\ref{fig:cv})
for the largest  $L$ suggests a continuous transition in the Ising-2D
universality class.
\begin{figure}
\includegraphics[angle=-90,width=7cm]{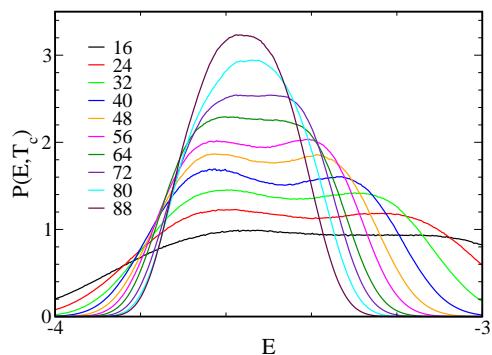}
\caption{(Color online) Probability distribution $P(E,T_c)$ of the
energy  per site at $T_c(L)$  for increasing linear $L$, from $L=16$ to
$L=88$, from bottom to top. $T_c(L)$ is   the temperature  where $C_v$
is  maximum. A  double-peak is visible  for $L\lesssim60$, and
disappears for $L=72$ and $L=80$.}
\label{fig:P(E)}
\end{figure}
To ascertain the continuous nature of the  transition, we computed the
probability distribution $P(E,T=T_c)$ of the energy  per site, obtained
from the density  of states $g(E)$ (Fig.~\ref{fig:P(E)}).  Here $T_c(L)$
is defined as the temperature where  the specific heat is maximum.
Rather surprisingly, from  small to intermediate lattice sizes,
$P(E,T=T_c)$ shows the bimodal structure characteristic of  a
first-order transition.  However, this feature disappears smoothly upon
increasing the system size, and a single  peak finally emerges for
$L\geq72$.
\begin{figure}
% JC: LaTeX se plaint trop: j'ai reduit la taille des figures mais la
% taille d'origine avait une bonne tete sur le .dvi
\subfigure[\label{fig:chicol}]{\includegraphics[angle=-90,width=6.5cm]{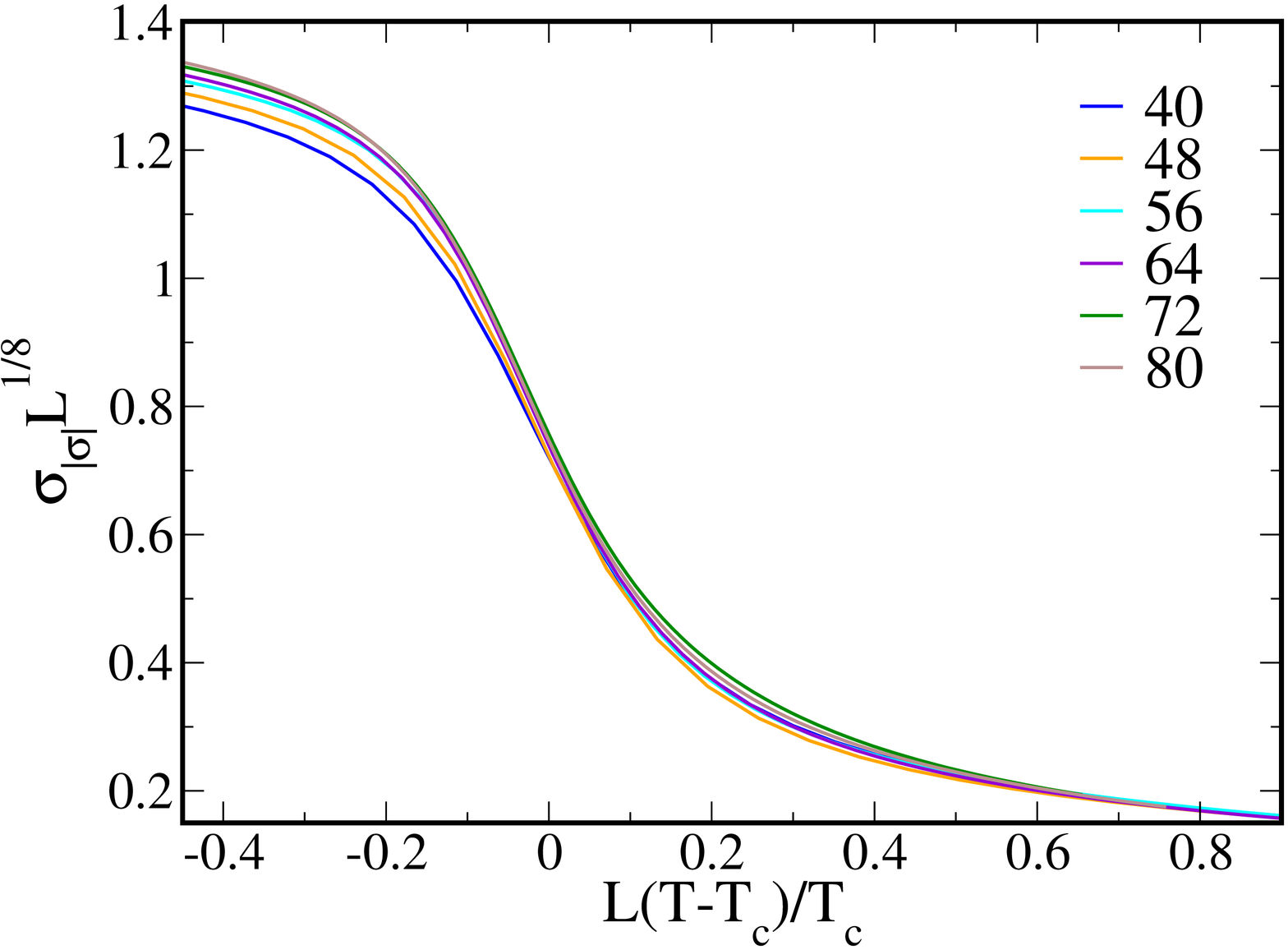}}\\
\subfigure[\label{fig:chisusc}]{\includegraphics[angle=-90,width=6.5cm]{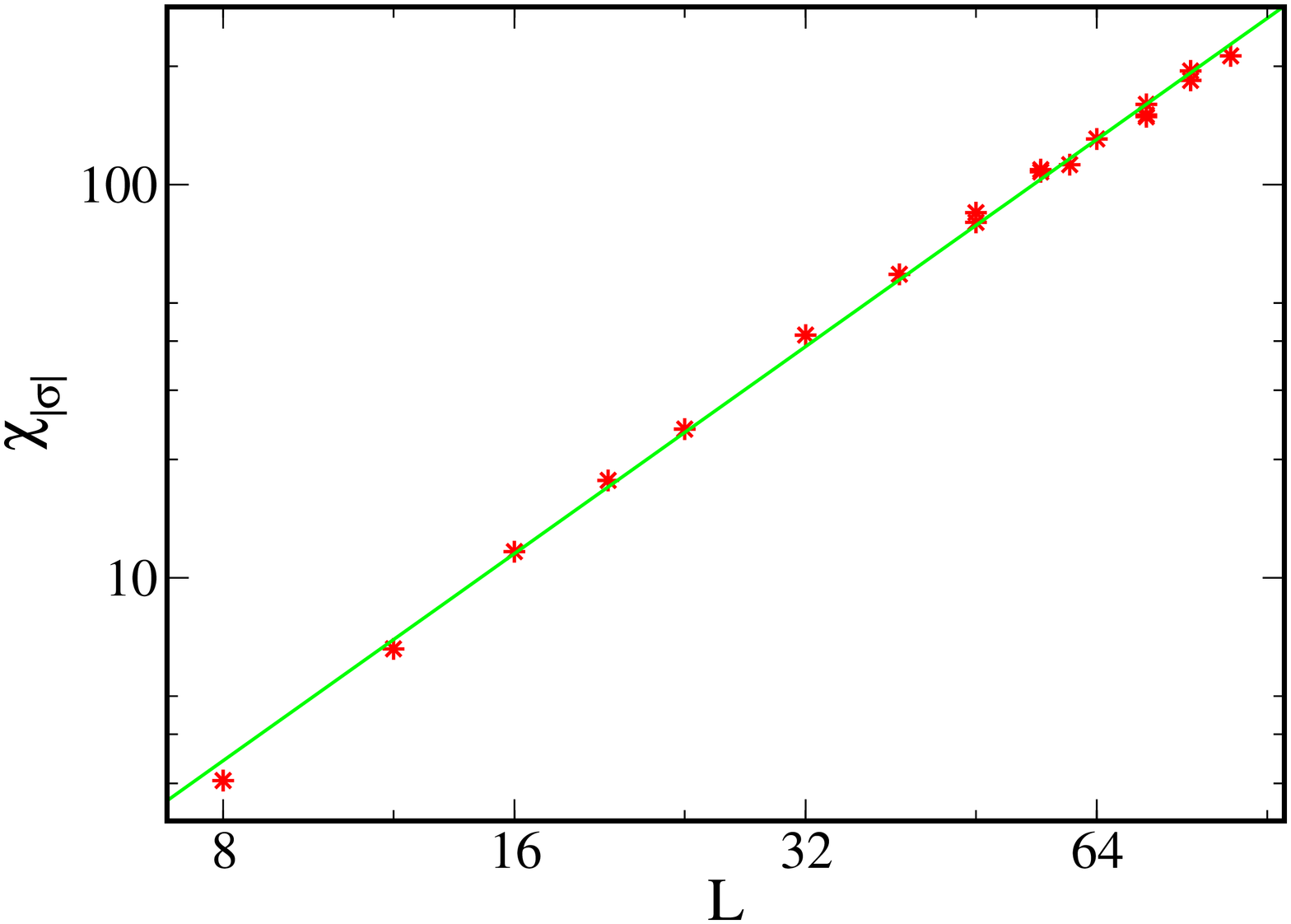}}\\
\subfigure[\label{fig:tc}]{\includegraphics[angle=-90,width=6.5cm]{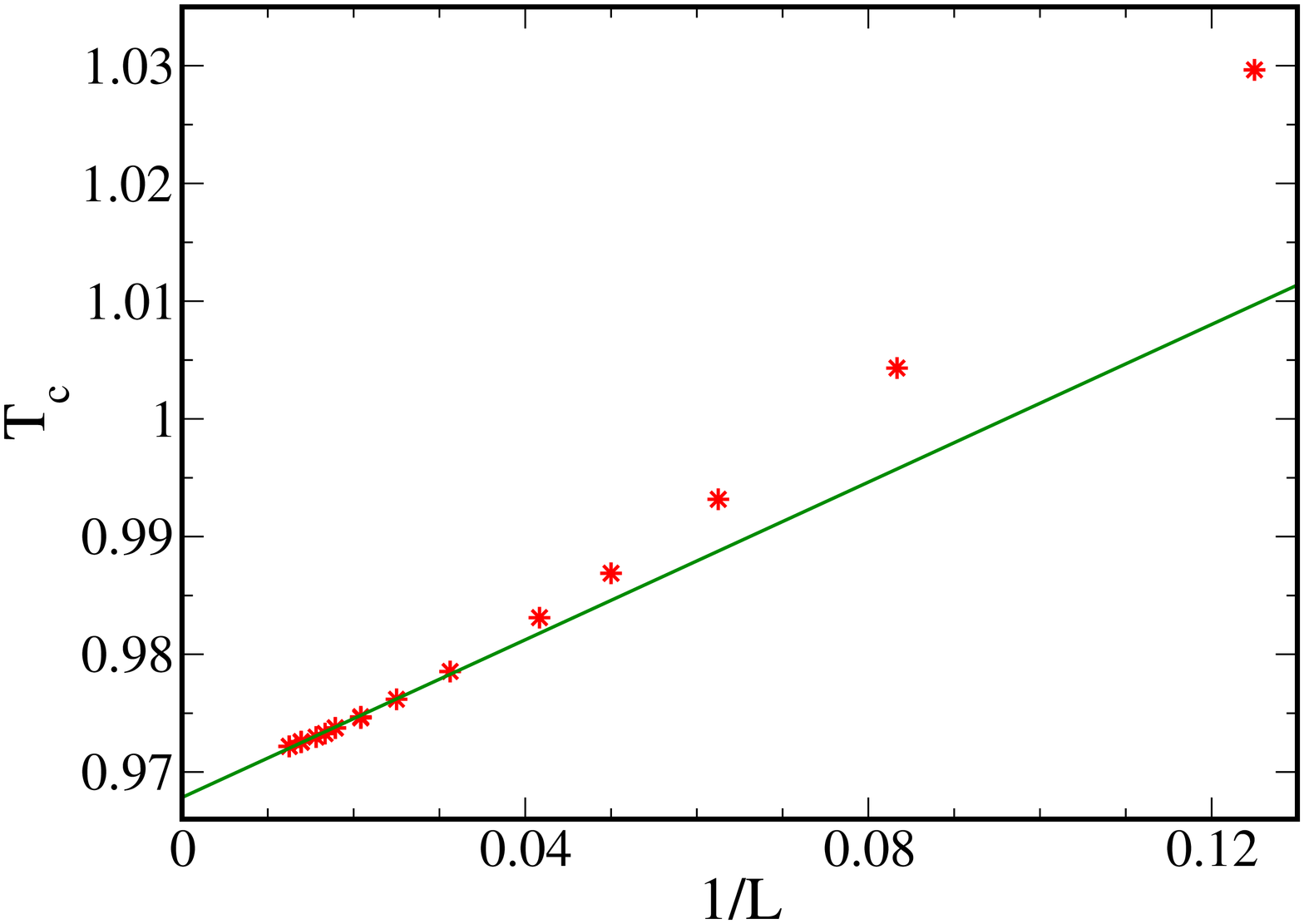}}\\
\subfigure[\label{fig:bind}]{\includegraphics[angle=-90,width=6.5cm]{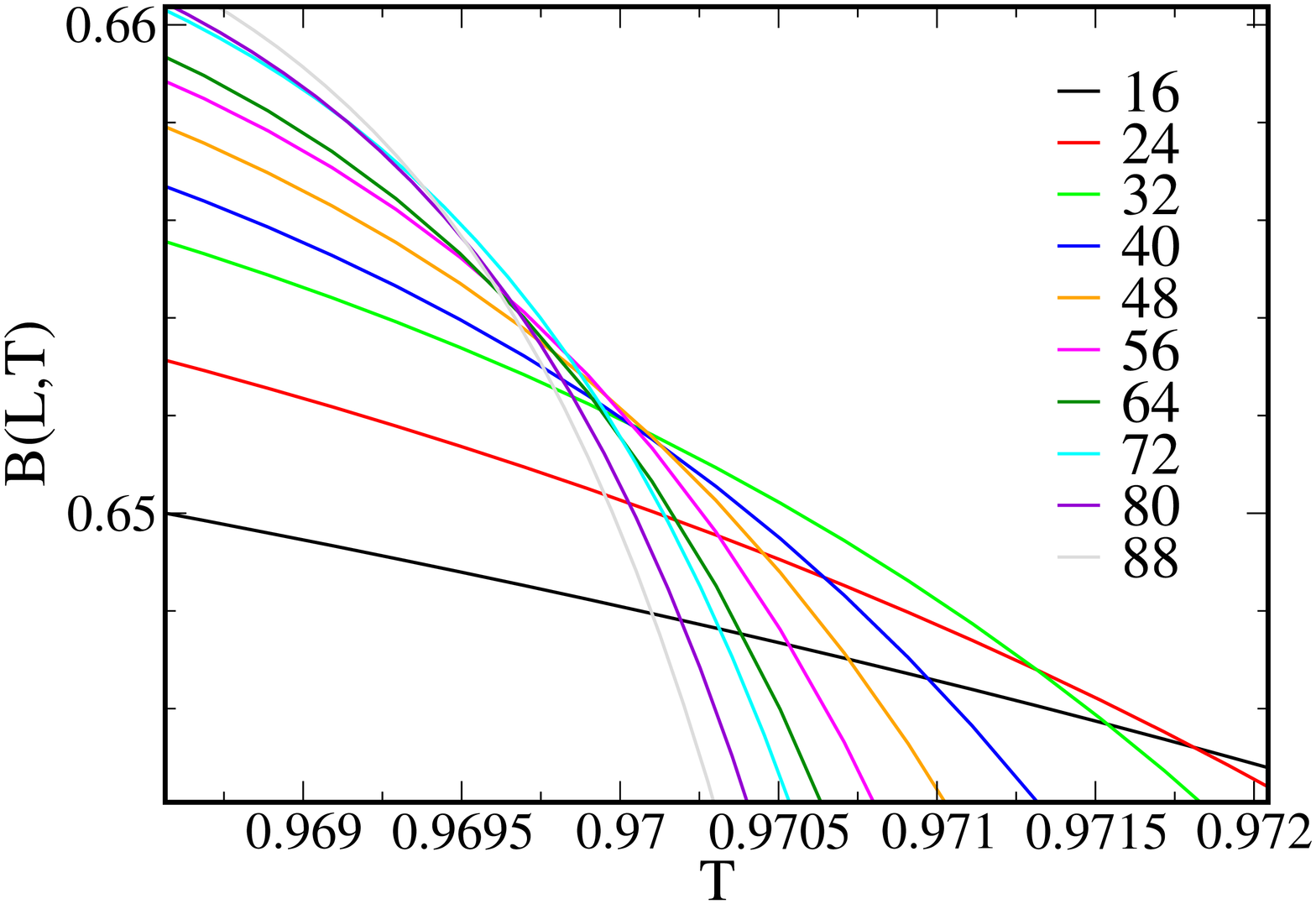}}
\caption{(Color online) Finite-size scaling at, or   near, the
transition temperature.  a) Rescaled chirality (Eq.~\ref{eq:chirality})
versus rescaled  temperature  for $40\leq L\leq80$.  b) Log-log plot of
the maximum of the  Ising susceptibility {\it vs.} $L$. c) Scaling  of
the maximum of the specific heat {\it vs.} $1/L$ (log-linear plot).  d)
Evolution of  the Binder cumulant $B(L,T)$ with     temperature, from
$L=16$ to $L=88$, from bottom to top.}
\label{fig:chirality}
\end{figure}

Hence we claim  that the phase transition  is {\it continuous} indeed,
in the Ising-2D universality class, although  the scaling and  energy
distribution  at  moderate sizes  is misleading.  This peculiar
finite-size  behavior evidences a large but finite length scale whose
exact nature has  not been elucidated so far.  We conjecture that it may
be related  to the proximity, in some parameter space, to  a tricritical
point   where  the   transition becomes discontinuous.  Further
arguments in support to this claim     will be provided       in
Sections~\ref{subsec:alterations}, \ref{sec:potts} and
\ref{sec:diluted}.

\subsection{Ising order parameter}
The ordering of the chirality variables $\sigma_i$ is probed by the
following Ising order parameter:
\begin{equation}
\sigma=\left<\frac{1}{N} \left| \sum_i \sigma_i \right|\right>,
\label{eq:chirality}
\end{equation}
where $N$ is the number of sites.  With the results above in mind we
analyze the finite-size  effects on the chirality using the scaling law
$L^{\beta/\nu}\sigma=f\left(L^{1/\nu}\left(\frac{T-T_c}{T_c}\right)\right)$
with $\beta=1/8$ and $\nu=1$. Fig.~\ref{fig:chicol} shows a data
collapse of the Ising order parameter for $40\leq L\leq88$, in agreement
with the  2D-Ising critical scenario ($\beta=1/8$ and $\nu=1$).
Figure~\ref{fig:chisusc} shows the scaling of the maximum of the Ising
susceptibility $\chi_\sigma$ versus $L$.   The  slope
$\gamma/\nu=1.76\pm0.02$  is  very close     to the expected   value of
7/4. We also plot the  temperature $T_c(L)$ of the the maximum of the
specific heat at size $L$ (Fig.~\ref{fig:tc}), showing the expected
asymptotic scaling $T_c(L)\sim T_\infty+A/L^{1/\nu}$ with $\nu=1$.
Finally,  we computed  the  fourth-order Binder cumulant
\begin{equation}
B(L,T)=1-\frac{1}{3}\frac{\langle \sigma^4 \rangle}{\langle \sigma^2
\rangle^2}
\end{equation}
which  shows  a characteristic $L$-independent  crossing at $T_c\simeq
0.97$ (Fig.~\ref{fig:bind}).   Note however  that    the value of the
cumulant at the crossing $B(L=72,T_{\textrm{crossing}})\simeq  0.65$
remains  larger  than the universal value of the 2D Ising model
(0.6107).  This discrepancy possibly originates from the fact that $L$
is not significantly larger than the crossover length-scale beyond which
the two peaks in  $P(E,T_c)$  merge (Sec.~\ref{sec:energy_distrib}),
making it uneasy to obtain a reliable estimate of the fourth moment of
the distribution of chiralities. We also point out that similar
discrepancies  were observed in the simulation of other emergent Ising
systems with continuous degrees of freedom.~\cite{weber03,numerics_FFXY}
\subsection{$\mathbb{Z}_2$ vortices in the Ising-$\mathbb{R}P^3$ model}
We now turn to  the identification of  $\mathbb{Z}_2$ vortices in  the
model defined in Eq.~\ref{eq:energy_vv2}.  The better-known case of an
$SO(3)$  ground state manifold  (ferromagnetically frozen chiralities)
was detailed  in  Section.~\ref{sec:fixed-chir-limit} Once  the chiral
degrees of freedom are relaxed, the ground  state manifold is enlarged
to  $O(3)=\mathbb{Z}_2\times SO(3)$  and  the definition  of vorticity
enclosed in lattice loops requires some  caution.  Let us consider two
sites $i$ and  $j$ and their associated  elements of $O(3)$, $M_i$ and
$M_j$. As long  as $M_i$ and $M_j$  have the same chirality,  no extra
difficulty   arises.  However,  if    they  have opposite  chiralities
(determinants),  then   the  mere  existence   of a    continuous path
connecting the  two elements in $SO(3)$ is  ill-posed, since they each
belong to a different $SO(3)$  sector of $O(3)$. Hence the computation
of the  circulation $\Omega(\mathcal{L})$ makes   sense only for loops
$\mathcal{L}$   enclosed  in   a  domain  of    uniform  chirality. In
particular, the computation    of the $\Z_2$ vorticity  on  plaquettes
located in the bulk  of uniform domains follows  the lines detailed in
Section~\ref{sec:fixed-chir-limit} without modification.

The case of non-uniform plaquettes,  sitting on a chiral domain  wall,
may   be addressed  in an   indirect way.  We  consider  a closed loop
$\mathcal{L}$ that i) only  visits sites with chirality $\sigma_i=+1$,
ii) encloses a domain of opposite chirality $\sigma_i=-1$.  i) ensures
that $\Omega(\mathcal{L})$ is well defined.  On the other hand one can
always define unambiguously  $N_v(\mathcal{L})$, the number  of vortex
cores  on uniform plaquettes   inside $\mathcal{L}$. If  the chirality
were              uniform       inside      $\mathcal{L}$         then
$\Omega(\mathcal{L})=(-1)^{N_v(\mathcal{L})}$ would  hold.    However,
when $\mathcal{L}$ encloses a domain with reversed chirality, an extra
contribution   arises    on the right    hand    side, coming from the
non-uniform  plaquettes sitting on  the  domain  wall,  which are  not
accounted       for          by      $(-1)^{N_v(\mathcal{L})}$.  Since
$\Omega(\mathcal{L})=\pm1$     we       obtain    $\Omega(\mathcal{L})
=\epsilon_\text{w}             (-1)^{N_v(\mathcal{L})}$          where
$\epsilon_\text{w}=\pm1$ acts as the  topological charge of the chiral
domain   wall. As a  result,  the present  workaround yields the total
$\Z_2$ charge carried by the chiral domain wall,  which is always well
defined.

\begin{figure}
\subfigure[\label{fig:corvor}]{\includegraphics[width=7cm]{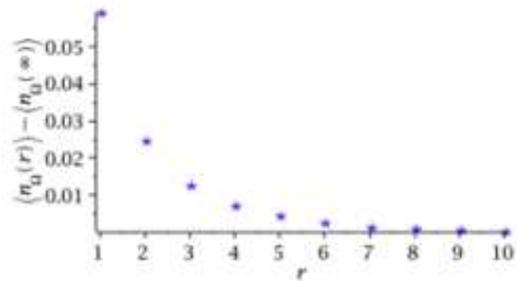}}\\
\subfigure[\label{fig:vortemp}]{\includegraphics[width=5cm,angle=-90]{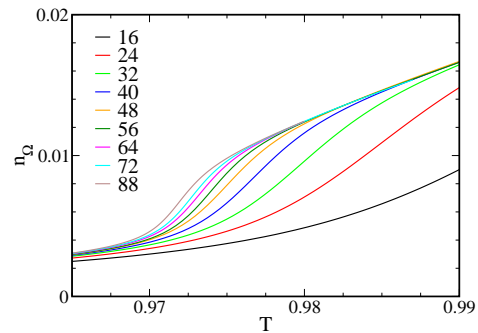}}
\subfigure[\label{fig:comp}]{\includegraphics[width=5cm,angle=-90]{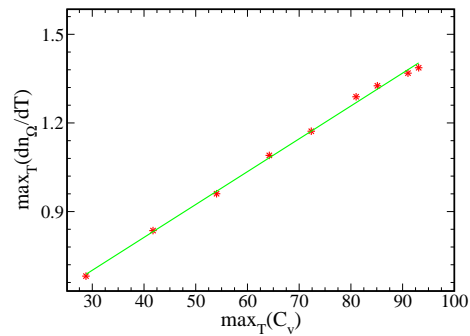}}
\caption{(Color online) a) Excess density of vortices {\it vs.} distance
to the domain wall at $T=1.3$ (see text).   b) Vortex density  in the
bulk of chiral domains {\it vs.} temperature $T$, from $L=16$ to $L=88$,
from bottom       to   top. c) Maximum of
$\left(\frac{dn_\Omega}{dT}\right) $ versus specific heat maximum ${\rm
max_T(C_v)}$ for $16 \leq L\leq88$. }
\label{fig:vortex-chirality}
\end{figure}

Figure~\ref{fig:confprop}  shows      a  typical   configuration  near
$T_c$. Again,  vortices in the bulk  of Ising domains are indicated by
(red) bullets A thorough  study of typical configurations reveals that
most of these vortex cores are actually ``paired''  with a nearby {\it
charged}        domain      wall          (not    represented       in
Fig.~\ref{fig:confprop}).   Once  the   charge   of  the   walls    is
appropriately  accounted  for,   the  total   measured   charge  is  1
indeed. Note  that such vortex/charged-wall  pairs feature  two $\Z_2$
charges, but involve the creation of only  one vortex core. Hence they
are energetically favored compared to genuine pairs of $\Z_2$ vortices
in the bulk of chiral domains. This is evidenced, for instance, by the
proliferation   of vortices   at  lower   temperatures   in the   full
Ising-$\mathbb{R}P^3$ than in the  $\mathbb{R}P^3$ model  (compare the
upper panel of Fig.\ref{fig:chi1} with Fig.\ref{fig:vortemp}).

To quantify the pairing effect  of $\Z_2$ vortices with charged walls,
we impose a domain  wall by splitting in the  system in two parts with
frozen but  opposite   chiralities (periodic boundary   conditions are
used).  Figure~\ref{fig:corvor} shows  the excess density of  vortices
near the domain wall,  compared  to the density in   the bulk  of  the
domains,  as   a  function of   the  distance   to  the interface  for
$T=1.3$.  In     agreement     with  the       trend observed       in
Fig.~\ref{fig:confprop}, vortex/charged-wall pairs are clearly favored
compared to vortex/vortex pairs in  the bulk. Moreover, this effect is
robust: the excess density of vortices is sizable  in a wide range of
temperatures.   Overall we anticipate  that   the same mechanism  will
prevail near more  complex interfaces, such as  those obtained in  the
full Ising-$\mathbb{R}P^3$ model at equilibrium.

For   completeness,  we computed the   core  vortex density $n_\Omega$
(defined  as the number of vortex  cores on uniform plaquettes divided
by the number  of such plaquettes) as   a function of  temperature for
different lattice sizes: the  increase  in the  vortex density at  the
transition temperature (Fig.~\ref{fig:vortemp}) scales with the system
size. This     is   apparent    on the     associated   susceptibility
$dn_{\Omega}/dT$ whose maximum scales with $L$ like the maximum of the
specific  heat (Fig.~\ref{fig:comp}). This is to  be compared with the
non-critical  behavior of the vortex   density in the  $\mathbb{R}P^3$
model   discussed    in      Section~\ref{sec:fixed-chir-limit}  where
$dn_{\Omega}/dT$ remains finite at all temperatures.

\subsection{Spatial correlations of discrete and continuous fluctuations}
%vectors and chiralities}
\label{sec:corr-betw-vect}

\begin{figure}
\subfigure[\label{fig:confprop}]{\includegraphics[width=7.5cm]{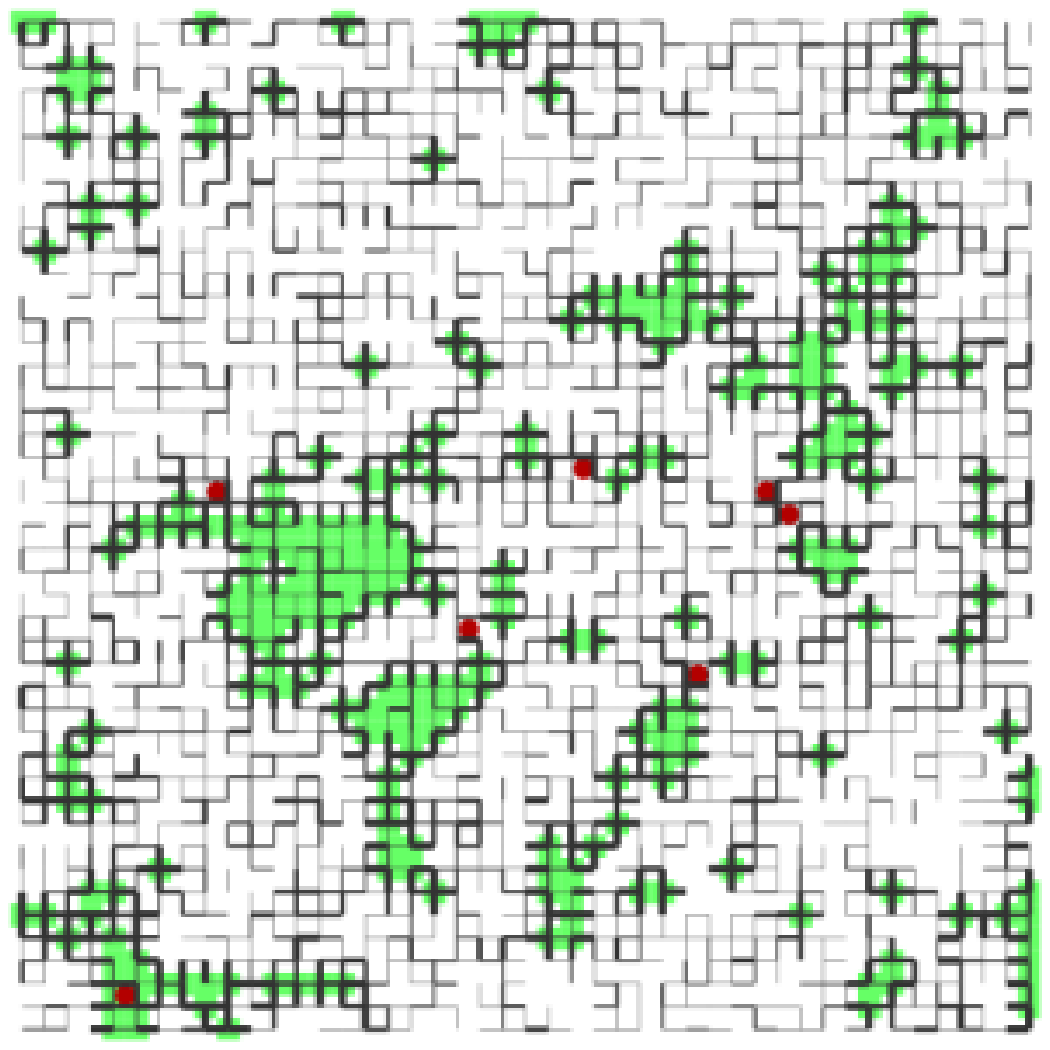}}
\subfigure[\label{fig:confsl}]{\includegraphics[width=7.5cm]{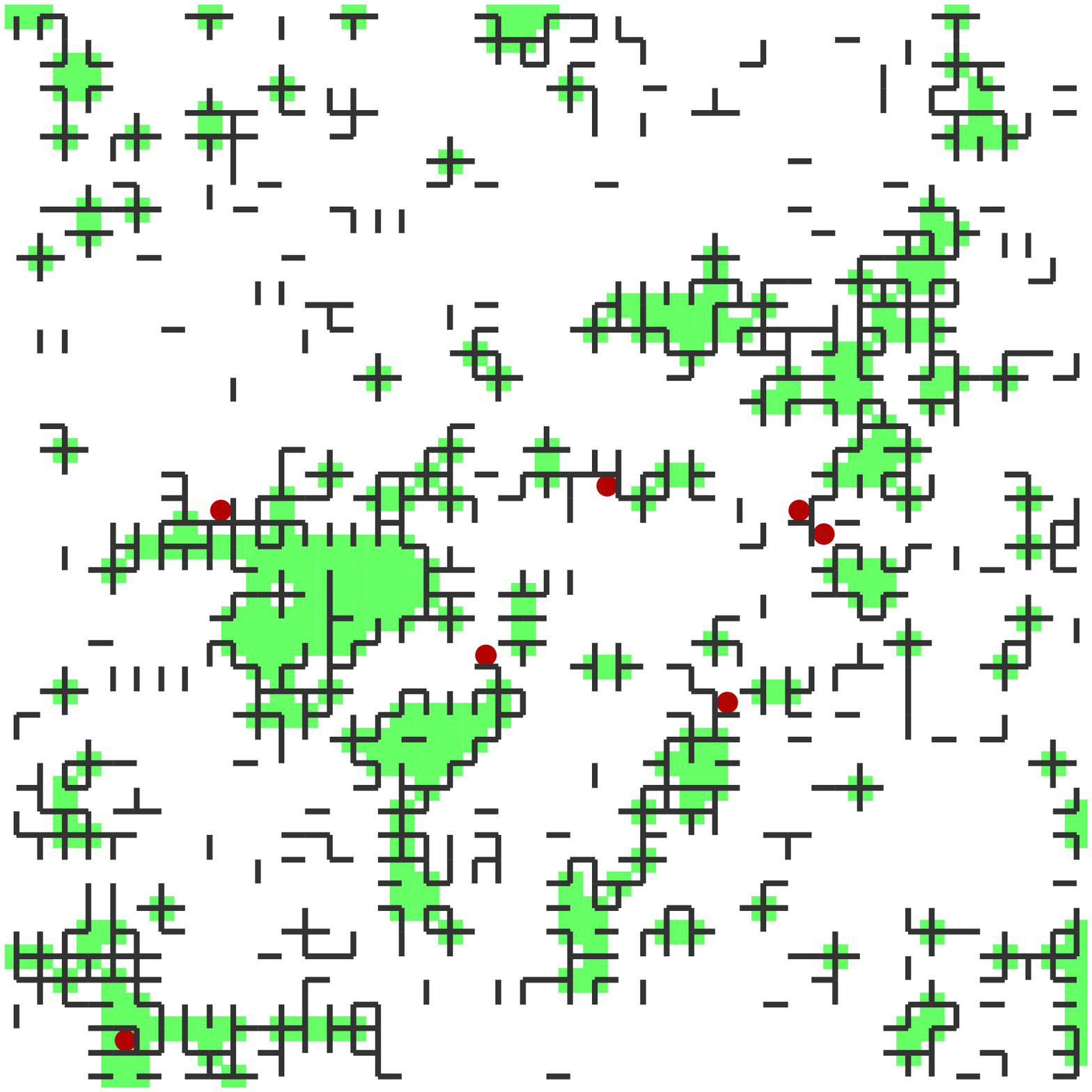}}
\caption{(Color online) Thermalized configuration at $T\simeq T_c$ of a
system  of  size  $L=72$  (only a   $45\times45$   square  snippet  is
shown). Red bullets indicate elementary square plaquettes that i) host
a   $\Z_2$    vortex  core    and  ii)  have     all  four  $\sigma_i$
equal.  Plaquettes with homogeneous   up (resp.  down) chiralities are
represented  by green   (resp white)  squares. a) the   width of bonds
$(i,j)$ is proportional  to $B_{ij}$ (see   text). b) Bonds are  drawn
only if $B_{ij}>1/2$.}
\label{fig:config}
\end{figure}
Figure~\ref{fig:confprop} also reveals  that the bonds standing across
chiral domain  walls   ($\sigma_i\sigma_j=-1$)   are also  the    most
``disordered'', or thickest, ones. This strong  correlation is all the
more     apparent   in   Figure~\ref{fig:confsl},    where   only  the
highest-energy,  most  disordered, bonds, are represented,  defined as
$B_{ij}>1/2$: they     are rather   scarce  in   the   bulk  of  Ising
domains.  Hence   the two kinds   of  fluctuations (associated  to the
discrete     Ising  variables   and to    the  continuous  4D-vectors,
respectively) are both  localized  close to  Ising domain  walls. As a
result,  the  energy barrier  for the formation  of  a  domain wall is
considerably  lowered compared   to  the pure  Ising   model (with all
4D-vectors   frozen   in  a  ferromagnetic   configuration).   This is
evidenced by the low  transition temperature $T_c=0.97$ for the  Ising
transition  observed  in  the   present Ising-$\mathbb{R}P^3$   model,
compared to $T_c=2.269$ for the Ising model in 2D.
\subsection{Small modifications of the Ising-$\mathbb{R}P^3$ model and
tuning of the nature of the phase transition}\label{subsec:alterations}
In view of the peculiar finite-size, first-order-like, behavior of the
energy distribution shown  in Figure~\ref{fig:P(E)}, we argue that the
Ising-$\mathbb{R}P^3$ model could be near  a tricritical point in some
parameter  space.  This is  consistent with the   nature  of the phase
transitions observed in a  number of related classical frustrated spin
models  with    similar    ``content'',   {\it   i.e.}     spin-waves,
$\Z_2$-vortices                      and                    Ising-like
chiralities.~\cite{domenge05,domenge08,momoi99}

In support of our claim, we present two distortions of the original
Hamiltonian~(\ref{eq:energy_vv2}) which preserve the ground state
symmetry, hence the nature of the excitations above, and show that they
undergo a first-order phase transition.

\begin{figure}
\subfigure[\label{fig:cv75}]{\includegraphics[angle=-90,width=7cm]{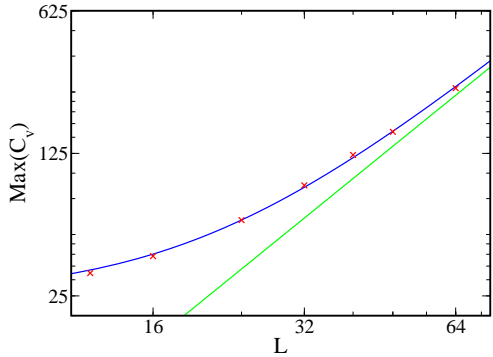}}\\
\subfigure[\label{fig:suscchi75}]{\includegraphics[angle=-90,width=7cm]{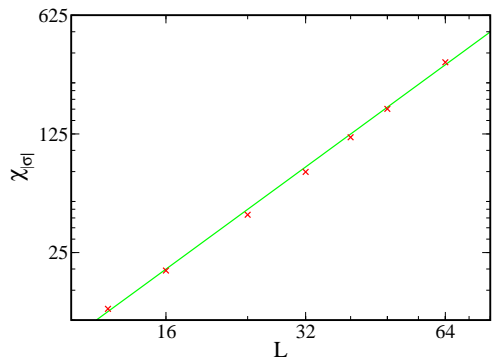}}
\caption{(Color online) a) Scaling as $L^2$ of the maximum of the
specific heat and b) of the chiral susceptibility  with increasing
system size $L$ for $a=1.75$ in Eq.~\ref{eq:a}.  }
\label{fig:model175}
\end{figure}

As it turns out, a simple change in the bond energy
\begin{equation}
E_{ij}=-\sigma_i\sigma_j(4(\vec v_i\cdot\vec v_j)^2-a)
\label{eq:a}
\end{equation}
from $a=1$ in the original Hamiltonian~(\ref{eq:energy_vv2}) to
$a=1.75$, is sufficient to  drive the   order-disorder transition of the
Ising variables towards first order. This    can   be seen     in
Figure~\ref{fig:model175}, where both the maximum of the specific heat
(Fig.~\ref{fig:cv75}) and  that    of   the chiral susceptibility
(Fig.~\ref{fig:suscchi75})       scale as $\sim L^2$.  Furthermore,
contrary to the Ising-$\mathbb{R}P^3$ case, the energy probability
distribution $P(E,T=T_c)$ remains bimodal for all lattice sizes, with  a
minimum that gets more and more pronounced upon increasing the system
size (not shown).

Increasing $a$ from $a=1$ in~(\ref{eq:a}) clearly decreases the energy
gap for a chirality flip $\sigma_i\to-\sigma_i$. However, the associated
modification of the entropy balance between the ordered and disordered
phase is less obvious.

Entropic effects are more explicit and better controlled in the
following family  of continuous spin models on 2D lattices:
\begin{equation}
E_{ij}=-\sum_{<i,j>}\left( \frac{1+\vec S_i\cdot\vec S_j}{2} \right)^p.
\label{eq:spinp}
\end{equation}
%
%where $\vec S_i$ are D-component spins.
Indeed, it was shown that for large enough $p$ ($p\gtrsim100$ for $XY$
spins~\cite{dsrs84}   and  $p\geq16$    for    Heisenberg
spins~\cite{bgh02,es02})  these systems undergo a  phase transition of
the liquid-gas type.  Obviously, tuning $p$ does not change the
energy scale ($E_{ij}\in[0,1]$), but increasing $p$ gradually pushes the
entropy toward  the highest  energies.

\begin{figure}
\includegraphics[angle=-90,width=7cm]{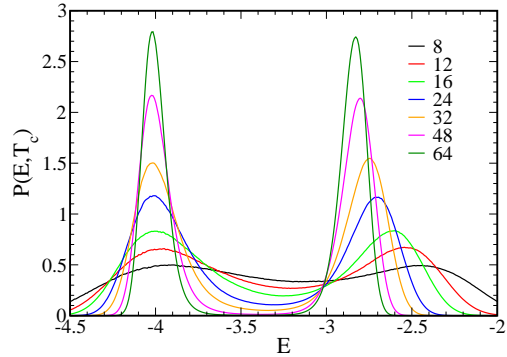}
\caption{(Color online) Energy distributions at the transition for
model~(\ref{eq:p}).}
\label{fig:model_quad}
\end{figure}

Hence we tweak the Ising-$\mathbb{R}P^3$ model in a similar way, with:
\begin{equation}
\label{eq:p}
E_{ij}=-\sigma_i\sigma_j(4(\vec v_i\cdot\vec v_j)^{2p}-1).
\end{equation}
Once again, Monte Carlo simulations of the distorted model~(\ref{eq:p})
shows the signatures of a first-order transition, for $p$ as small as
$p=2$ (Fig.~\ref{fig:model_quad}).

Overall the previous two models illustrate our claim on the proximity of
the   Ising transition in  the Ising-$\mathbb{R}P^3$ model with a
first-order transition.

\section{Entropy of $\sigma_i\sigma_j$ bonds -- Potts model analogy}
\label{sec:potts}
\begin{figure}
\includegraphics[width=5cm,angle=0]{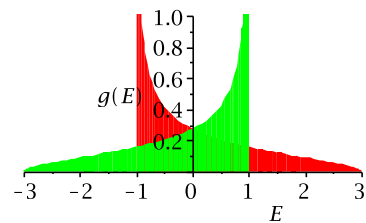}\\
\caption{(Color online) Density of state $g^\epsilon(E)$ for a single
bond in model~(\ref{eq:energy_vv2}). The two colors correspond to
$\epsilon=\sigma_i\sigma_j=1$ (green) and $\epsilon=-1$ (red). The
lowest energy ($E=-3$ per bond) is attained when $(\vec v_i\cdot\vec
v_j)^2=1$ and $\sigma_i \sigma_j=1$. 
}
\label{fig:g(E)}
\end{figure}
Here we take another route and  propose a simple qualitative analogy  to
explain how the coupling of  chiralities $\sigma_i$  to the continuous
degrees of freedom $\vec v_i$   drives  the Ising  transition  in the
Ising-$\mathbb{R}P^3$ model close to a first-order transition.

To that end we compute the density of states $g(E)$ for a single bond
$(i,j)$.  It has two contributions $g(E)=g^+(E)+g^-(E)$ depending on the
value of $\epsilon=\sigma_i\sigma_j$:
\begin{eqnarray}
g^{\epsilon}(E)&\sim& \int_{S^3} d^3\vec v_i \int_{S^3} d^3\vec v_j
\delta\left(E+\epsilon(4(\vec v_i\cdot\vec v_j)^2-1)\right) \nonumber \\
&\sim&\int_{S^3} d^3\vec v_i
\delta\left(E+\epsilon(4(v_i^0)^2-1)\right)
\end{eqnarray}
where rotational invariance was used to fix $\vec v_j=[1,0,0,0]$.
Explicit evaluation of the integral gives
\begin{equation}
g^\epsilon(-3\leq \epsilon E < 1)=\frac{1}{2\pi}\sqrt{\frac{3+\epsilon
E}{1-\epsilon E}}.
\label{eq:pottsp1}
\end{equation}
Figure~\ref{fig:g(E)} shows the evolution of $g^{\pm}(E)$ with $E$.
Interestingly, the density  of state is remarkably small in the vicinity
of the (ferromagnetic) ground state value ($\sigma_i\sigma_j=1$, $\vec
v_i\cdot\vec v_j=1$,  and $E=-3$). However, if the two trihedra have
opposite chiralities, the lowest energy state is attained for orthogonal
vectors $\vec v_i\cdot\vec v_j=0$, with energy $E=-1$, and the
associated density of states diverges.

This strong entropy unbalance between ordered ($\epsilon=1$) and
disordered bonds ($\epsilon=-1$) is similar to  that of the well-known
$q-$state Potts models.~\cite{baxter73}  In this  model, each ``spin''
$\sigma$ can take $q$ different colors, and the interaction energy is
$E=0$ for neighboring sites $(i,j)$ with the same color
$\sigma_i=\sigma_j$, and $E=1$ otherwise, with the resulting density of
states $$ g(0\leq E\leq 1)=q\delta(E)+q(q-1)\delta(E-1). $$ Hence
``disordered'' configurations  indeed carry more weight than the
ferromagnetic ground state.
This entropy unbalance is seen to increase  with $q$ and in 2D it is
known to eventually drive the order-disorder transition towards first
order for $q>4$.~\cite{baxter73}

A similar feature is obtained upon distorting the Ising-$\mathbb{R}P^3$
model as in~(\ref{eq:p}). Indeed, the computation of the
density of states $g_p(E)$ for $p=2$ yields
\begin{equation}
g_{p=2}^\epsilon(-3\leq \epsilon E <
1)=\frac{1}{2\pi}\frac{\sqrt{2-\sqrt{1-\epsilon E}}}{(1-\epsilon
E)^{3/4} },
\label{eq:pottsp2}
\end{equation}
which has the same features as in Fig.~\ref{fig:g(E)} except that it
diverges as $\sim x^{-3/4}$ at $E=\pm1$, instead of $\sim x^{-1/2}$.
Hence, the distortion of the bond energy shown in~(\ref{eq:p})
essentially increases the entropy unbalance, which ultimately drives the
Ising transition towards first-order (Sec.~\ref{subsec:alterations}),
much in the same way as in the $q-$states Potts model.

To elaborate further on the proximity  to a first order transition, it
is  useful to recall some results from the real-space  renormalization
group (RG) treatment of the $q$-states Potts model.~\cite{nbrs79}
As noted by Nienhuis {\it et al.},~\cite{nbrs79} in 2D, conventional RG
approaches give accurate results in the critical regime ($q\leq4$) but
inexplicably fail to predict the crossover to a discontinuous transition
for $q\geq4$.  The authors proposed that it originates from the usual
coarse-graining procedure, by which a single Potts spin is assigned to a
finite region in real space, using a majority rule.  Intuitively, this
brutal substitution becomes physically questionable when there is no
clear majority spin in the domain, a situation that is likely to occur
when the number of colors $q$ is large enough. In particular, it yields the
possibility of a ferromagnetic effective interaction between such
(artificially) polarized super-cells, even if the microscopic spins are
disordered. Hence, conventional coarse-graining overestimates the
tendency to ferromagnetic order.

In Ref.~\cite{nbrs79}  it is argued  that disordered  regions interact
only weakly with their neighbors, hence they are better coarse-grained
as a  vacancy  (missing Pott   spin).  In  support to  this  intuitive
picture, it was  shown that, once the parameter  space of the original
Potts Hamiltonian    is enlarged to  include   the fugacity   of these
vacancies, the  real-space RG treatment of the  Potts model is able to
detect the  crossover of  the   order-disorder transition, seen  as  a
liquid-gaz transition of the vacancies.

In the following section we discuss a simplified version of the
Ising-$\mathbb{R}P^3$ model where discrete variables $t_i$ are
introduced. The latter play a role  similar  to that of the vacancies in
Ref.~\onlinecite{nbrs79}.
\section{Effective diluted Ising model} \label{sec:diluted}

In  this section  we propose  a   simplified  model, with  strong
analogies       to      the     Ising-$\mathbb{R}P^3$
model~(\ref{eq:energy_vv2}), that captures the spatial correlations
evidenced in Fig.\ref{fig:confsl} and where the entropy unbalance
discussed above for the Potts model, is at play.

We replace  the vector degrees  of freedom with {\it discrete} variables
$t_i=0,1$, so that the energy becomes:
\begin{equation}
E=-\sum_{\langle i,j\rangle} \sigma_i \sigma_j (4 t_i t_j -1)
+D(T)\sum_i t_i.
\label{eq:chi-t}
\end{equation}
We introduce a temperature-dependent  ``chemical potential'' $D(T)$ to
tune  $\langle t_i\rangle$   (hence $\langle   t_i t_j\rangle$).   The
relation with   the    original  model~(\ref{eq:energy_vv2})  can   be
understood as follows.  A site  with $t_i=1$ represents a vector which
is  collinear  (or almost  collinear) with  the  ``majority''  of  its
neighbors. On the other hand, a  site with $t_i=0$ represents a vector
which  is perpendicular (or almost perpendicular)  to  the majority of
its neighbors.  The fact that the  vector-vector correlation length is
significantly  larger than one lattice  spacing at the temperatures of
interest justifies   that, locally,  the vectors have   a well-defined
local orientation.  Then,   we simply replace  $(\vec v_i   \cdot \vec
v_j)^2$ by $t_i t_j$.  Of  course, in the  original model, two vectors
$\vec v_i$ and $\vec v_j$ can be  simultaneously i) orthogonal to most
of  their neighbors ($t_i=t_j=0$)    and ii) parallel  to  each  other
($(\vec  v_i    \cdot \vec  v_j)^2=1$):  such  situations  are clearly
discarded      by    this    discretized   model.~\footnote{  In   the
Ising-$\mathbb{R}P^3$  model,   domain   walls  separate  regions with
$\sigma=+1$ from  those  with $\sigma=-1$.  Fig.~\ref{fig:g(E)}  shows
that ferromagnetic  configurations of the vectors   are favored in the
bulk   of    chirality   domains.  On   the   other  hand,  orthogonal
configurations  are  favored   on   bonds that   stand   across domain
walls.  Hence the lowest  energy configuration for  the  vectors is to
point in direction, say, $[1,0,0,0]$ in the $\sigma=+1$ domain, and in
some orthogonal direction,  say $[0,1,0,0]$ in the $\sigma=-1$ domain.
An   analog low-energy   configuration  in   the  discrete  model   of
Eq.~\ref{eq:chi-t} is obtained by setting  $t_i=1$ in the bulk of both
domains, and  $t_j=0$ in the vicinity of  the domain wall.} As a final
encouragement to study  the discrete model of  Eq.~\ref{eq:chi-t}), we
mention that its single bond density of state is qualitatively similar
to that of the original model (Fig.~\ref{fig:g(E)}): it has two ground
states at $E=-3$ ($\sigma_i=\sigma_j$ and $t_i=t_j=1$), and $2\times3$
excited   states  associated to   chiral  flip  at  $E=-1$ ($\sigma_i=
-\sigma_j$ and $t_i t_j=0$).

\begin{figure}
\includegraphics[width=7.5cm]{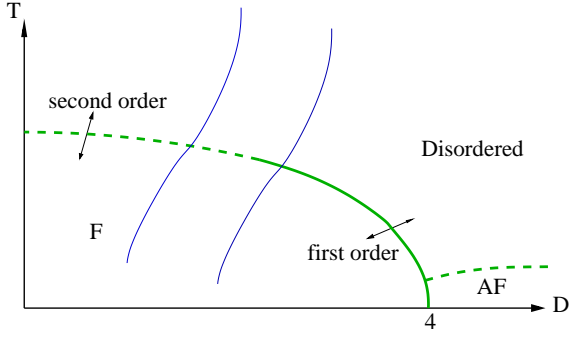}
\caption{(Color online) Schematic mean-field phase diagram of the
discrete model~(\ref{eq:chi-t}) in  the  $(T,D)$ plane.  The  two thin
lines are trajectories of our models (\ref{eq:p}), with $p=1,2$).}
\label{fig:cctt}
\end{figure}

This    model     closely   resembles   the     celebrated Blume-Capel
model,~\cite{blume-capel}  where an  Ising  transition  becomes  first
order when the concentration of ``holes'' (sites with $t_i=0$) becomes
large    enough.~\footnote{The    Hamiltonian of   the   Blume   Capel
model~\cite{blume-capel}        is      $H=-J\sum_{<i,j>}S_iS_j+\Delta
\sum_iS^2_i$, where $S_i$  denotes  a  spin-1 at   site $i$,  $J$  the
ferromagnetic  interaction and    $\Delta$  the  crystal-field.    The
Blume-Capel  model  and the  site-diluted  spin model~(\ref{eq:chi-t})
become   equivalent  once   the   $-1$  contribution  is   dropped  in
Eq.~\ref{eq:chi-t}   and  $D(T)=\Delta   -T\ln(2)$.}  Since  a  simple
mean-field approximation is   sufficient predict the first and  second
order transition lines of the Blume-Capel model  has (depending on the
crystal  field parameter $\Delta$), we  determine the mean-field phase
diagram of   Eq.~\ref{eq:chi-t},   using two   mean-field   parameters
$\langle t_i\rangle=t$ and $\langle
\sigma_i\rangle=\sigma$.  Fig.~\ref{fig:cctt} shows that it is composed
of   four   transition    lines,   and   one  obtains   four  distinct
order-to-disorder    transitions    depending  on   the    value    of
$D(T)$. Namely,  upon increasing  $D(T)$ we  find  i) a  second  order
ferromagnetic-paramagnetic transition  (large  negative  $D$),   ii) a
first order ferromagnetic-paramagnetic  transition, iii) a first order
ferromagnetic-antiferromagnetic transition  and  iv)  a second   order
antiferromagnetic-paramagnetic  transition.  Hence we conjecture  that
the   present     model also   has a  tricritical    point,  where the
ferromagnetic to paramagnetic  transition changes from second order to
first order.     This  approach  is    obviously too   crude   to   be
quantitatively accurate, but  we  can  still  make contact   with  the
Ising-$\mathbb{R}P^3$ model by adjusting the chemical potential $D(T)$
to enforce $\langle t_i t_j\rangle=\langle  (\vec v_i\cdot \vec v_j)^2
\rangle$   (the  right-hand    side is  computed      in the  original
Ising-$\mathbb{R}P^3$ model).   Upon  changing  the temperature,  this
model describes a curve  in  the $D-T$  plane  such as those shown  in
Fig.~\ref{fig:cctt}.  At   $T=\infty$,  $\langle  (\vec  v_i\cdot \vec
v_j)^2 \rangle=\frac{1}{4}$ and  one  can  show  that $D(T=\infty)=0$.
Further, upon decreasing the  temperature $\langle (\vec v_i\cdot \vec
v_j)^2 \rangle$  decreases     and  $D(T)$   decreases, until      the
ferromagnetic phase of the discrete model is reached.

Overall this supports our claim  that the unusual finite-size behavior
of  the Ising-$\mathbb{R}P^3$ model  (Fig.~\ref{fig:P(E)}), as well as
the   proximity     to    a    first    order      chiral   transition
(Section~\ref{subsec:alterations})    originates     from    a  nearby
tricritical  point  in parameter   space.  Further, the present  study
suggests that the  nature and origin of  the tricritical  point can be
understood, at least qualitatively, from an effective Blume-Capel like
model.
\section{Effective Ising model with multiple-spin interactions}
\label{sec:HT}
In this section we detail   a  more  quantitative approach  to the
Ising-$\mathbb{R}P^3$ model, in which the vector spins $\vec v_i$ are
integrated out perturbatively in $\beta=1/T$, in order to  derive  an
effective Ising model  for   the  chirality degrees of freedom.  From
this standpoint, multiple spin interactions between chiralities are
directly responsible for the ``proximity'' to a first order transition.
\subsection{Integrating out the vector spins}
Because  the order parameter of  the transition is  the chirality, and
because the vector spins never  order at $T>0$, it  is natural to look
for an effective   model    involving {\it only}   the    chiralities.
Moreover, we have shown that the Ising domain walls are accompanied by
short distance rearrangements of the 4D-vectors, and that it is a very
important aspect of the energetics of the system. It is thus natural to
expect that a high temperature expansion for the vector spins (which
captures short-distance correlations) will be semi-quantitatively valid.

Formally, the integration over  the  4D-vectors leads to the following
energy $E_{\rm eff }$ for a configuration $\{\sigma_i\}$ of the chiralities:
\begin{equation}
E_{\rm eff }(\{\sigma_i\})= -T\ln\left( \left< e^{-\beta
E(\{\sigma_i,\vec v_i\})} \right> \right),
\label{eq:Eeff}
\end{equation}
where $E(\{\sigma_i,\vec v_i\})$ is given by Eq.~\ref{eq:energy_vv2},
and $\left< \cdots \right>$ is the \mbox{$T=\infty$} average for each vector
spin $\vec v_i\in S^3$ (uniform measure on $S^3\times\cdots \times
S^3$).

In the following, we derive the effective interaction of the chiralities
by  expanding Eq~(\ref{eq:Eeff}) in powers of $\beta=1/T$ up to order
$\beta^8$.
\subsection{$1/T$ expansion}
\begin{eqnarray}
\left< e^{-\beta E(\{\sigma_i,\vec v_i\})} \right> &=&
\sum_{n=0}^{\infty} \frac{(-\beta)^n}{n!} \nonumber\\
\times\sum_{\langle i_1,j_1\rangle} \sum_{\langle i_2,j_2\rangle} \cdots
\sum_{\langle i_n,j_n\rangle} &&	
%\nonumber\\&&
\left<E_{i_1j_1} E_{i_2j_2}\cdots E_{i_nj_n} \right> \nonumber\\
&&\times\sigma_{i_1} \sigma_{j_1}\cdots \sigma_{i_n}\sigma_{j_n}
\label{eq:exp_av}
\end{eqnarray}
with
\begin{eqnarray}
E_{ij}&=&1-4(\vec v_i \cdot \vec v_j)^2
\end{eqnarray}
As $\left<E_{ij}\right>=0$, expanding     the  logarithm of
Eq.~\ref{eq:Eeff}   in  powers of $\beta$  yields the following cumulant
expansion:
\begin{align}
&E_{\rm eff}(\{\sigma_i\})=-T\left[ \frac{\beta^2}{2!} \sum_{\langle
i_1,j_1\rangle}\sum_{\langle i_2,j_2\rangle}
C^2_{i_1j_1i_2j_2}\sigma_{i_1} \sigma_{j_1}\sigma_{i_2} \sigma_{j_2}
\right. \nonumber \\ &-\frac{\beta^3}{3!}\sum_{\langle
i_1,j_1\rangle}\sum_{\langle i_2,j_2\rangle}\sum_{\langle
i_3,,j_3\rangle}
C^3_{i_1j_1i_2j_2i_3j_3}\sigma_{i_1}\sigma_{j_1}\sigma_{i_2}
\sigma_{j_2}\sigma_{i_3} \sigma_{j_3} \nonumber \\
&\left.+\frac{\beta^4}{4!}\sum_{\langle
i_1,j_1\rangle}\cdots\sum_{\langle i_4,j_4\rangle} C^4_{i_1j_1\cdots
i_4j_4}\sigma_{i_1} \sigma_{j_1}\cdots\sigma_{i_4} \sigma_{j_4}
+\mathcal{O}(\beta^5)\right]
\label{eq:EeffC}
\end{align}
with the cumulants
\begin{eqnarray}
C^2_{i_1j_1i_2j_2}&=& \left<E_{i_1j_1} E_{i_2j_2}\right>\\
C^3_{i_1j_1i_2j_2i_3j_3}&=&\left<E_{i_1j_1}
E_{i_2j_2}E_{i_3j_3}\right>\\ C^4_{i_1j_1\cdots
i_4j_4}&=&\left<E_{i_1j_1} \cdots E_{i_4j_4}\right>\nonumber\\
&&-\left<E_{i_1j_1} E_{i_2j_2}\right>\left<E_{i_3j_3}
E_{i_4j_4}\right>\nonumber\\ &&-\left<E_{i_1j_1}
E_{i_3j_3}\right>\left<E_{i_2j_2} E_{i_4j_4}\right>\nonumber\\
&&-\left<E_{i_1j_1} E_{i_4j_4}\right>\left<E_{i_2j_2} E_{i_3j_3}\right>
\end{eqnarray}
As usual in series expansion, the cumulant $C^n$ is non-zero only if the
graph defined by the $n$ bonds $(i_1,j_1),\cdots,(i_n,j_n)$ is
connected.  Moreover, rotational invariance ensures that only graphs
that are one-particle-irreducible contribute.  For this reason,
$C^2_{i_1j_1i_2j_2}$ is non-zero only when the two bonds coincide,
resulting in a constant contribution (independent of the $\sigma_i$) to
$E_{\rm eff}$.  At the next order, and for the same reason, the only
non-zero  $C^3$ come from graphs where the three bonds coincide (and
$C^3_{121212}=1$). This generates an effective first neighbor Ising
interaction proportional to $1/T^2$:
\begin{equation}
E_{\rm eff}=-\frac{1}{6T^2}\sum_{\langle i,j\rangle} \sigma_i \sigma_j.
\end{equation}

$C^4$  only provides a constant  contribution to the effective energy.
$C^5$   and  $C^6$ terms introduce new   two-chirality interactions,
between first, second and third neighbors. Moreover, a $C^6$ term gives
the first interaction with more than two chiralities, namely:
\begin{equation}
-\frac{1}{9 T^5}\sum_{\langle i,j,k,l\rangle} \sigma_i \sigma_j \sigma_k
\sigma_l,
\end{equation}
where the sum runs over square plaquettes.

At  this order in $\beta$, the  Ising-$\mathbb{R}P_3$ model appears as
an Ising model  with two and   four spin interactions.   The effect of
such multiple-spin interactions has   been  studied for the   3D Ising
model,  where  an additional four  spin interaction,  if   it is large
enough, can make the transition first order.~\cite{mkjf81} This can be
understood,  at  least qualitatively, from a   very  simple mean field
point of  view.   Indeed, $p-$spin  interactions will  translate  into
terms of the order of $m^p$ in  the Landau free  energy ($m$ being the
order   parameter). Hence it  is  clear  that tuning  the strength  of
multiple  ($\geq 4$)  spin interactions  can reshape   the free energy
landscape, and drive the transition from second to first-order.

However, various   approaches  predicted   that the   simplest  4-spin
interactions were not enough to obtain a first-order transition in two
dimensions.~\cite{og73,o74,gm77}     To check  these   predictions  we
performed  Monte Carlo simulations of  a simple Ising model with first
neighbor  coupling,     supplemented  with  a     4-spins    plaquette
interaction.\footnote{Although the  two-spin interactions appearing in
$E_{\rm eff}$ up  to order $T^{-6}$ are not  strictly limited to first
neighbors   (see   the   graphs    in  Table~\ref{tab:graphs}).}   The
Hamiltonian reads
\begin{equation}
H=-J\sum_{<ij>}\sigma_i\sigma_j-K\sum_{\square}\sigma_i\sigma_j\sigma_k\sigma_l,
\end{equation}
with  $J,K>0$. Using finite-size scaling analysis, we find that the
transition remains of second-order on the whole range $0\leq K/J\leq10$.

This lead us to continue  the  high-temperature expansion up to  order
$\beta^8$, where a   multi-spin interaction involving  6 chiralities is
generated (with new  2- and 4-interaction terms).  The   effective
Hamiltonian becomes quite complicated
and   therefore  we resort   to a  simple mean-field calculation.
\begin{table}

\begin{ruledtabular}

\begin{tabular}{cc}
\begin{tabular}{|cc|}

\hline
$g$&$\mathcal{N}_g$\\
\hline
\hline
\rule{18pt}{0pt}
\psline(-0.3,0.1)(0.3,0.1)
\psline[linearc=0.5]%
       (-0.3,0.1)(0,0.25)(0.3,0.1)
\psline[linearc=0.5]%
       (-0.3,0.1)(0,-0.05)(0.3,0.1)
\rule{14pt}{0pt}
&$2$
\\

\hline
\hline
\psline(-0.3,0.1)(0.3,0.1)
\psline[linearc=0.3]%
       (-0.3,0.1)(0,0.3)(0.3,0.1)
\psline[linearc=0.5]%
       (-0.3,0.1)(0,0.2)(0.3,0.1)
\psline[linearc=0.5]%
       (-0.3,0.1)(0,0.0)(0.3,0.1)
\psline[linearc=0.3]%
       (-0.3,0.1)(0,-0.1)(0.3,0.1)
&$2$
\\

\hline
\psline(-.3,0.4)(0.3,0.4)
\psline(-.3,-0.2)(0.3,-0.2)
\psline(-.3,-0.2)(-.3,0.4)
\psline[linearc=0.5]%
       (0.3,-0.2)(0.22,0.1)(0.3,0.4)
\psline[linearc=0.5]%
       (0.3,-0.2)(0.38,0.1)(0.3,0.4)
&$240$
\rule[-9pt]{0pt}{23pt}
\\

\hline\hline
\psline(-.3,0.4)(0.3,0.4)
\psline(-.3,0.4)(-0.3,-0.2)
\psline[linearc=0.5]%
       (-0.3,-0.2)(-0,-0.28)(0.3,-0.2)
\psline[linearc=0.5]%
       (-0.3,-0.2)(-0,-0.12)(0.3,-0.2)
\psline[linearc=0.5]%
       (0.3,-0.2)(0.22,0.1)(0.3,0.4)
\psline[linearc=0.5]%
       (0.3,-0.2)(0.38,0.1)(0.3,0.4)
&$720 $
\rule[-9pt]{0pt}{23pt}
\\

\hline
\psline(-.3,0.4)(0.3,0.4)
\psline(-.3,-0.2)(0.3,-0.2)
\psline[linearc=0.5]%
       (0.3,-0.2)(0.22,0.1)(0.3,0.4)
\psline[linearc=0.5]%
       (0.3,-0.2)(0.38,0.1)(0.3,0.4)
\psline[linearc=0.5]%
       (-0.3,-0.2)(-0.22,0.1)(-0.3,0.4)
\psline[linearc=0.5]%
       (-0.3,-0.2)(-0.38,0.1)(-0.3,0.4)
&$360$
\rule[-9pt]{0pt}{23pt}
\\

\hline\hline
\psline(-0.3,0.1)(0.3,0.1)
\psline[linearc=0.5]%
       (-0.3,0.1)(0,0.2)(0.3,0.1)
\psline[linearc=0.5]%
       (-0.3,0.1)(0,0.0)(0.3,0.1)
\psline[linearc=0.3]%
       (-0.3,0.1)(0,0.3)(0.3,0.1)
\psline[linearc=0.3]%
       (-0.3,0.1)(0,-0.1)(0.3,0.1)
\psline[linearc=0.3]%
       (-0.3,0.1)(0,-0.32)(0.3,0.1)
\psline[linearc=0.3]%
       (-0.3,0.1)(0,0.52)(0.3,0.1)
&$2$
\rule[-5pt]{0pt}{17pt}
\\

\hline
\psline[linearc=0.6]%
       (0.3,-0.2)(0.35,0.1)(0.3,0.4)
\psline[linearc=0.6]%
       (0.3,-0.2)(0.25,0.1)(0.3,0.4)
\psline[linearc=0.4]%
       (0.3,-0.2)(0.47,0.1)(0.3,0.4)
\psline[linearc=0.4]%
       (0.3,-0.2)(0.13,0.1)(0.3,0.4)
\psline(-.3,0.4)(0.3,0.4)
\psline(-.3,0.4)(-0.3,-0.2)
\psline(.3,-0.2)(-0.3,-0.2)
&$840 $
\rule[-9pt]{0pt}{23pt}
\\

\hline
\psline(-.3,0.4)(0.3,0.4)
\psline(-.3,0.4)(-0.3,-0.2)
\psline[linearc=0.5]%
       (-0.3,-0.2)(-0,-0.28)(0.3,-0.2)
\psline[linearc=0.5]%
       (-0.3,-0.2)(-0,-0.12)(0.3,-0.2)
\psline(0.3,-0.2)(0.3,0.4)
\psline[linearc=0.5]%
       (0.3,-0.2)(0.15,0.1)(0.3,0.4)
\psline[linearc=0.5]%
       (0.3,-0.2)(0.45,0.1)(0.3,0.4)
&$3360 $
\rule[-9pt]{0pt}{23pt}
\\

\hline
\psline(-.3,0.4)(0.3,0.4)
\psline(-.3,-0.2)(0.3,-0.2)
\psline(0.3,-0.2)(0.3,0.4)
\psline[linearc=0.5]%
       (0.3,-0.2)(0.15,0.1)(0.3,0.4)
\psline[linearc=0.5]%
       (0.3,-0.2)(0.45,0.1)(0.3,0.4)
\psline[linearc=0.5]%
       (-0.3,-0.2)(-0.22,0.1)(-0.3,0.4)
\psline[linearc=0.5]%
       (-0.3,-0.2)(-0.38,0.1)(-0.3,0.4)
&$1680$
\rule[-9pt]{0pt}{23pt}
\\

\hline
\psline(-.3,0.4)(0.3,0.4)
\psline[linearc=0.5]%
       (0.3,-0.2)(0.22,0.1)(0.3,0.4)
\psline[linearc=0.5]%
       (0.3,-0.2)(0.38,0.1)(0.3,0.4)
\psline[linearc=0.5]%
       (-0.3,-0.2)(-0.22,0.1)(-0.3,0.4)
\psline[linearc=0.5]%
       (-0.3,-0.2)(-0.38,0.1)(-0.3,0.4)
\psline[linearc=0.5]%
       (-0.3,-0.2)(-0,-0.28)(0.3,-0.2)
\psline[linearc=0.5]%
       (-0.3,-0.2)(-0,-0.12)(0.3,-0.2)
&$2520$
\rule[-9pt]{0pt}{23pt}
\\

\hline
\psline(0,-0.2)(0.6,-0.2)
\psline(0.6,-0.2)(0.6,0.4)
\psline(0,0.4)(0.6,0.4)
\psline(-.6,0.4)(0,0.4)
\psline(-.6,0.4)(-0.6,-0.2)
\psline(0,-0.2)(0,0.4)
\psline(-0.6,-0.2)(0,-0.2)
&$10080$
\rule[-9pt]{0pt}{23pt}
\\

\hline
\psline(0,-0.2)(0.6,-0.2)
\psline(0.6,-0.2)(0.6,0.4)
\psline(0,0.4)(0.6,0.4)
\psline(-.6,0.4)(0,0.4)
\psline(-.6,0.4)(-0.6,-0.2)
\psline[linearc=0.6]%
       (-0.6,-0.2)(-0.3,-0.12)(0,-0.2)
\psline[linearc=0.6]%
       (-0.6,-0.2)(-0.3,-0.28)(0,-0.2)
&$30240$
\rule[-9pt]{0pt}{23pt}
\\

\hline\hline
\psline[linearc=0.6]%
       (0.3,-0.2)(0.35,0.1)(0.3,0.4)
\psline[linearc=0.6]%
       (0.3,-0.2)(0.25,0.1)(0.3,0.4)
\psline[linearc=0.4]%
       (0.3,-0.2)(0.47,0.1)(0.3,0.4)
\psline[linearc=0.4]%
       (0.3,-0.2)(0.13,0.1)(0.3,0.4)
\psline(-.3,0.4)(0.3,0.4)
\psline(-.3,0.4)(-0.3,-0.2)
\psline[linearc=0.6]%
       (-0.3,-0.2)(0,-0.12)(0.3,-0.2)
\psline[linearc=0.6]%
       (-0.3,-0.2)(0,-0.28)(0.3,-0.2)
&$6720$
\rule[-9pt]{0pt}{23pt}
\\

\hline
\psline[linearc=0.6]%
       (0.3,-0.2)(0.35,0.1)(0.3,0.4)
\psline[linearc=0.6]%
       (0.3,-0.2)(0.25,0.1)(0.3,0.4)
\psline[linearc=0.4]%
       (0.3,-0.2)(0.47,0.1)(0.3,0.4)
\psline[linearc=0.4]%
       (0.3,-0.2)(0.13,0.1)(0.3,0.4)
\psline(-.3,0.4)(0.3,0.4)
\psline[linearc=0.5]%
       (-0.3,-0.2)(-0.22,0.1)(-0.3,0.4)
\psline[linearc=0.5]%
       (-0.3,-0.2)(-0.38,0.1)(-0.3,0.4)
\psline(-.3,-0.2)(0.3,-0.2)
&$3360$
\rule[-9pt]{0pt}{23pt}
\\

\hline
\psline[linearc=0.5]%
       (-0.3,-0.2)(-0.22,0.1)(-0.3,0.4)
\psline[linearc=0.5]%
       (-0.3,-0.2)(-0.38,0.1)(-0.3,0.4)
\psline(-.3,0.4)(0.3,0.4)
\psline[linearc=0.5]%
       (-0.3,-0.2)(-0,-0.28)(0.3,-0.2)
\psline[linearc=0.5]%
       (-0.3,-0.2)(-0,-0.12)(0.3,-0.2)
\psline(0.3,-0.2)(0.3,0.4)
\psline[linearc=0.5]%
       (0.3,-0.2)(0.15,0.1)(0.3,0.4)
\psline[linearc=0.5]%
       (0.3,-0.2)(0.45,0.1)(0.3,0.4)
&$13440$
\rule[-9pt]{0pt}{23pt}
\\

\hline
\psline(-.3,0.4)(-0.3,-0.2)
\psline[linearc=0.5]%
       (-0.3,-0.2)(0,-0.28)(0.3,-0.2)
\psline[linearc=0.5]%
       (-0.3,-0.2)(0,-0.12)(0.3,-0.2)
\psline[linearc=0.5]%
       (-0.3,0.4)(0,0.48)(0.3,0.4)
\psline[linearc=0.5]%
       (-0.3,0.4)(0,0.32)(0.3,0.4)
\psline(0.3,-0.2)(0.3,0.4)
\psline[linearc=0.5]%
       (0.3,-0.2)(0.15,0.1)(0.3,0.4)
\psline[linearc=0.5]%
       (0.3,-0.2)(0.45,0.1)(0.3,0.4)
&$6720$
\rule[-9pt]{0pt}{24pt}
\\

\hline
\psline(0,-0.2)(0.6,-0.2)
\psline(0,0.4)(0.6,0.4)
\psline(0,-0.2)(0,0.4)
\psline(0.6,-0.2)(0.6,0.4)
\psline(-.6,0.4)(0,0.4)
\psline(-.6,0.4)(-0.6,-0.2)
\psline[linearc=0.6]%
       (-0.6,-0.2)(-0.3,-0.12)(0,-0.2)
\psline[linearc=0.6]%
       (-0.6,-0.2)(-0.3,-0.28)(0,-0.2)
&$161280$
\rule[-9pt]{0pt}{23pt}
\\

\hline
\psline(0,-0.2)(0.6,-0.2)
\psline(0,0.4)(0.6,0.4)
\psline(0,-0.2)(0,0.4)
\psline(-.6,0.4)(0,0.4)
\psline(-.6,0.4)(-0.6,-0.2)
\psline(-0.6,-0.2)(0,-0.2)
\psline[linearc=0.5]%
       (0.6,-0.2)(0.52,0.1)(0.6,0.4)
\psline[linearc=0.5]%
       (0.6,-0.2)(0.68,0.1)(0.6,0.4)
&$80640$
\rule[-9pt]{0pt}{23pt}
\\
\hline

\hline
\psline(0,-0.2)(0.6,-0.2)
\psline(0,0.4)(0.6,0.4)
\psline(-.6,0.4)(0,0.4)
\psline(-.6,0.4)(-0.6,-0.2)
\psline[linearc=0.6]%
       (-0.6,-0.2)(-0.3,-0.12)(0,-0.2)
\psline[linearc=0.6]%
       (-0.6,-0.2)(-0.3,-0.28)(0,-0.2)
\psline[linearc=0.5]%
       (0.6,-0.2)(0.52,0.1)(0.6,0.4)
\psline[linearc=0.5]%
       (0.6,-0.2)(0.68,0.1)(0.6,0.4)
&$120960$
\rule[-9pt]{0pt}{23pt}
\\

\hline

\hline
\psline(0,-0.2)(0.6,-0.2)
\psline(0,0.4)(0.6,0.4)
\psline(-.6,0.4)(0,0.4)
\psline(-.6,-0.2)(0,-0.2)
\psline[linearc=0.5]%
       (-0.6,-0.2)(-0.52,0.1)(-0.6,0.4)
\psline[linearc=0.5]%
       (-0.6,-0.2)(-0.68,0.1)(-0.6,0.4)
\psline[linearc=0.5]%
       (0.6,-0.2)(0.52,0.1)(0.6,0.4)
\psline[linearc=0.5]%
       (0.6,-0.2)(0.68,0.1)(0.6,0.4)
&$60480$
\rule[-9pt]{0pt}{23pt}
\\

\hline
\psline(0,-0.2)(-0.6,-0.2)
\psline(0,0.4)(0.6,0.4)
\psline(-.6,0.4)(0,0.4)
\psline(-.6,0.4)(-0.6,-0.2)
\psline[linearc=0.6]%
       (0.6,-0.2)(0.3,-0.12)(0,-0.2)
\psline[linearc=0.6]%
       (0.6,-0.2)(0.3,-0.28)(0,-0.2)
\psline[linearc=0.5]%
       (0.6,-0.2)(0.52,0.1)(0.6,0.4)
\psline[linearc=0.5]%
       (0.6,-0.2)(0.68,0.1)(0.6,0.4)
&$120960$
\rule[-9pt]{0pt}{23pt}
\\
\hline

\end{tabular}

\begin{tabular}{|cc|}

\hline
\rule{18pt}{0pt}
\psline(-0.3,0.1)(0.3,0.1)
\psline[linearc=0.5]%
       (-0.3,0.1)(0,0.2)(0.3,0.1)
\psline[linearc=0.5]%
       (-0.3,0.1)(0,0.0)(0.3,0.1)
\psline[linearc=0.3]%
       (-0.3,0.1)(0,0.3)(0.3,0.1)
\psline[linearc=0.3]%
       (-0.3,0.1)(0,-0.1)(0.3,0.1)
\psline[linearc=0.3]%
       (-0.3,0.1)(0,-0.32)(0.3,0.1)
\psline[linearc=0.3]%
       (-0.3,0.1)(0,0.52)(0.3,0.1)
\psline[linearc=0.28]%
       (-0.3,0.1)(0,-0.6)(0.3,0.1)
\psline[linearc=0.28]%
       (-0.3,0.1)(0,0.8)(0.3,0.1)
\rule{14pt}{0pt}
&$2$
\rule[-7pt]{0pt}{19pt}
\\

\hline
\psline[linearc=0.6]%
       (0.3,-0.2)(0.35,0.1)(0.3,0.4)
\psline[linearc=0.6]%
       (0.3,-0.2)(0.25,0.1)(0.3,0.4)
\psline[linearc=0.4]%
       (0.3,-0.2)(0.47,0.1)(0.3,0.4)
\psline[linearc=0.4]%
       (0.3,-0.2)(0.13,0.1)(0.3,0.4)
\psline[linearc=0.3]%
       (0.3,-0.2)(0.6,0.1)(0.3,0.4)
\psline[linearc=0.3]%
       (0.3,-0.2)(0,0.1)(0.3,0.4)
\psline(-.3,0.4)(0.3,0.4)
\psline(-.3,0.4)(-0.3,-0.2)
\psline(-0.3,-0.2)(0.3,-0.2)
&$2016$
\rule[-9pt]{0pt}{23pt}
\\

\hline
\psline(0.3,-0.2)(0.3,0.4)
\psline[linearc=0.3]%
       (0.3,-0.2)(0.5,0.1)(0.3,0.4)
\psline[linearc=0.5]%
       (0.3,-0.2)(0.2,0.1)(0.3,0.4)
\psline[linearc=0.5]%
       (0.3,-0.2)(0.4,0.1)(0.3,0.4)
\psline[linearc=0.3]%
       (0.3,-0.2)(0.1,0.1)(0.3,0.4)
\psline(-.3,-0.2)(-0.3,0.4)
\psline(-.3,0.4)(0.3,0.4)
\psline[linearc=0.6]%
       (-0.3,-0.2)(0,-0.12)(0.3,-0.2)
\psline[linearc=0.6]%
       (-0.3,-0.2)(0,-0.28)(0.3,-0.2)
&$12096$
\rule[-9pt]{0pt}{23pt}
\\

\hline
\psline(0.3,-0.2)(0.3,0.4)
\psline[linearc=0.3]%
       (0.3,-0.2)(0.5,0.1)(0.3,0.4)
\psline[linearc=0.5]%
       (0.3,-0.2)(0.2,0.1)(0.3,0.4)
\psline[linearc=0.5]%
       (0.3,-0.2)(0.4,0.1)(0.3,0.4)
\psline[linearc=0.3]%
       (0.3,-0.2)(0.1,0.1)(0.3,0.4)
\psline(-.3,0.4)(0.3,0.4)
\psline[linearc=0.5]%
       (-0.3,-0.2)(-0.22,0.1)(-0.3,0.4)
\psline[linearc=0.5]%
       (-0.3,-0.2)(-0.38,0.1)(-0.3,0.4)
\psline(-.3,-0.2)(0.3,-0.2)
&$6048$
\rule[-9pt]{0pt}{23pt}
\\

\hline
\psline[linearc=0.6]%
       (0.3,-0.2)(0.35,0.1)(0.3,0.4)
\psline[linearc=0.6]%
       (0.3,-0.2)(0.25,0.1)(0.3,0.4)
\psline[linearc=0.4]%
       (0.3,-0.2)(0.47,0.1)(0.3,0.4)
\psline[linearc=0.4]%
       (0.3,-0.2)(0.13,0.1)(0.3,0.4)
\psline(-.3,0.4)(0.3,0.4)
\psline(-.3,0.4)(-0.3,-0.2)
\psline(-.3,-0.2)(0.3,-0.2)
\psline[linearc=0.5]%
       (-0.3,-0.2)(0,-0.05)(0.3,-0.2)
\psline[linearc=0.5]%
       (-0.3,-0.2)(0,-0.35)(0.3,-0.2)
&$20160$
\rule[-10pt]{0pt}{24pt}
\\

\hline
\psline[linearc=0.6]%
       (0.3,-0.2)(0.35,0.1)(0.3,0.4)
\psline[linearc=0.6]%
       (0.3,-0.2)(0.25,0.1)(0.3,0.4)
\psline[linearc=0.4]%
       (0.3,-0.2)(0.47,0.1)(0.3,0.4)
\psline[linearc=0.4]%
       (0.3,-0.2)(0.13,0.1)(0.3,0.4)
\psline(-.3,0.4)(0.3,0.4)
\psline(-.3,0.4)(-0.3,-0.2)
\psline(-.3,-0.2)(0.3,-0.2)
\psline[linearc=0.5]%
       (-0.3,-0.2)(-0.15,0.1)(-0.3,0.4)
\psline[linearc=0.5]%
       (-0.3,-0.2)(-0.45,0.1)(-0.3,0.4)
&$10080$
\rule[-9pt]{0pt}{23pt}
\\

\hline
\psline(-.3,-0.2)(0.3,-0.2)
\psline[linearc=0.5]%
       (0.3,-0.2)(0.15,0.1)(0.3,0.4)
\psline[linearc=0.5]%
       (0.3,-0.2)(0.45,0.1)(0.3,0.4)
\psline(.3,-0.2)(0.3,0.4)
\psline[linearc=0.5]%
       (-0.3,-0.2)(0,-0.05)(0.3,-0.2)
\psline[linearc=0.5]%
       (-0.3,-0.2)(0,-0.35)(0.3,-0.2)
\psline[linearc=0.5]%
       (-0.3,-0.2)(-0.22,0.1)(-0.3,0.4)
\psline[linearc=0.5]%
       (-0.3,-0.2)(-0.38,0.1)(-0.3,0.4)
\psline(.3,0.4)(-0.3,0.4)
&$40320$
\rule[-10pt]{0pt}{24pt}
\\

\hline

\psline[linearc=0.5]%
       (0.3,-0.2)(0.15,0.1)(0.3,0.4)
\psline[linearc=0.5]%
       (0.3,-0.2)(0.45,0.1)(0.3,0.4)
\psline(.3,-0.2)(0.3,0.4)
\psline[linearc=0.5]%
       (-0.3,-0.2)(-0.15,0.1)(-0.3,0.4)
\psline[linearc=0.5]%
       (-0.3,-0.2)(-0.45,0.1)(-0.3,0.4)
\psline(-.3,-0.2)(-0.3,0.4)
\psline(.3,0.4)(-0.3,0.4)
\psline[linearc=0.6]%
       (-0.3,-0.2)(0,-0.12)(0.3,-0.2)
\psline[linearc=0.6]%
       (-0.3,-0.2)(0,-0.28)(0.3,-0.2)

&$20160$
\rule[-9pt]{0pt}{23pt}
\\

\hline
\psline(-.3,-0.2)(0.3,-0.2)
\psline[linearc=0.5]%
       (-0.3,-0.2)(0,-0.05)(0.3,-0.2)
\psline[linearc=0.5]%
       (-0.3,-0.2)(0,-0.35)(0.3,-0.2)
\psline[linearc=0.5]%
       (-0.3,-0.2)(-0.22,0.1)(-0.3,0.4)
\psline[linearc=0.5]%
       (-0.3,-0.2)(-0.38,0.1)(-0.3,0.4)
\psline[linearc=0.6]%
       (-0.3,0.4)(0,0.32)(0.3,0.4)
\psline[linearc=0.6]%
       (-0.3,0.4)(0,0.48)(0.3,0.4)
\psline[linearc=0.5]%
       (0.3,-0.2)(0.22,0.1)(0.3,0.4)
\psline[linearc=0.5]%
       (0.3,-0.2)(0.38,0.1)(0.3,0.4)
&$30240$
\rule[-10pt]{0pt}{25pt}
\\

\hline
\psline[linearc=0.6]%
       (0.3,-0.2)(0.35,0.1)(0.3,0.4)
\psline[linearc=0.6]%
       (0.3,-0.2)(0.25,0.1)(0.3,0.4)
\psline[linearc=0.4]%
       (0.3,-0.2)(0.47,0.1)(0.3,0.4)
\psline[linearc=0.4]%
       (0.3,-0.2)(0.13,0.1)(0.3,0.4)
\psline(-.3,0.4)(-0.3,-0.2)
\psline[linearc=0.5]%
       (-0.3,-0.2)(-0,-0.28)(0.3,-0.2)
\psline[linearc=0.5]%
       (-0.3,-0.2)(-0,-0.12)(0.3,-0.2)
\psline[linearc=0.5]%
       (-0.3,0.4)(-0,0.48)(0.3,0.4)
\psline[linearc=0.5]%
       (-0.3,0.4)(-0,0.32)(0.3,0.4)
&$15120$
\rule[-9pt]{0pt}{24pt}
\\

\hline
\psline[linearc=0.6]%
       (0.3,-0.2)(0.35,0.1)(0.3,0.4)
\psline[linearc=0.6]%
       (0.3,-0.2)(0.25,0.1)(0.3,0.4)
\psline[linearc=0.4]%
       (0.3,-0.2)(0.47,0.1)(0.3,0.4)
\psline[linearc=0.4]%
       (0.3,-0.2)(0.13,0.1)(0.3,0.4)
\psline(-.3,0.4)(0.3,0.4)
\psline[linearc=0.5]%
       (-0.3,-.2)(-0,-0.28)(0.3,-0.2)
\psline[linearc=0.5]%
       (-0.3,-.2)(-0,-0.12)(0.3,-0.2)
\psline[linearc=0.5]%
       (-0.3,-0.2)(-0.38,0.1)(-0.3,0.4)
\psline[linearc=0.5]%
       (-0.3,-0.2)(-0.22,0.1)(-0.3,0.4)
&$30240$
\rule[-9pt]{0pt}{23pt}
\\

\hline
\psline[linearc=0.5]%
       (0,-0.2)(-0.15,0.1)(0,0.4)
\psline[linearc=0.5]%
       (0,-0.2)(0.15,0.1)(0,0.4)
\psline(0,-0.2)(0.6,-0.2)
\psline(0.6,-0.2)(0.6,0.4)
\psline(0,0.4)(0.6,0.4)
\psline(-.6,0.4)(0,0.4)
\psline(-.6,0.4)(-0.6,-0.2)
\psline(0,-0.2)(0,0.4)
\psline(-0.6,-0.2)(0,-0.2)
&$120960$
\rule[-9pt]{0pt}{23pt}
\\

\hline
\psline[linearc=0.5]%
       (0,-0.2)(-0.08,0.1)(0,0.4)
\psline[linearc=0.5]%
       (0,-0.2)(0.08,0.1)(0,0.4)
\psline[linearc=0.5]%
       (0.6,-0.2)(0.52,0.1)(0.6,0.4)
\psline[linearc=0.5]%
       (0.6,-0.2)(0.68,0.1)(0.6,0.4)
\psline(0,-0.2)(0.6,-0.2)
\psline(0,0.4)(0.6,0.4)
\psline(-.6,0.4)(0,0.4)
\psline(-.6,0.4)(-0.6,-0.2)
\psline(-0.6,-0.2)(0,-0.2)
&$362880$
\rule[-9pt]{0pt}{23pt}
\\

\hline
\psline[linearc=0.5]%
       (0,-0.2)(-0.08,0.1)(0,0.4)
\psline[linearc=0.5]%
       (0,-0.2)(0.08,0.1)(0,0.4)
\psline(0.6,-0.2)(0.6,0.4)
\psline(0,0.4)(0.6,0.4)
\psline(-.6,0.4)(0,0.4)
\psline(-.6,0.4)(-0.6,-0.2)
\psline(-0.6,-0.2)(0,-0.2)
\psline[linearc=0.6]%
       (0.6,-0.2)(0.3,-0.12)(0,-0.2)
\psline[linearc=0.6]%
       (0.6,-0.2)(0.3,-0.28)(0,-0.2)
&$725760$
\rule[-9pt]{0pt}{23pt}
\\

\hline
\psline[linearc=0.5]%
       (0,-0.2)(0.3,-0.05)(0.6,-0.2)
\psline[linearc=0.5]%
       (0,-0.2)(0.3,-0.35)(0.6,-0.2)
\psline(0,-0.2)(0.6,-0.2)
\psline(0.6,-0.2)(0.6,0.4)
\psline(0,0.4)(0.6,0.4)
\psline(-.6,0.4)(0,0.4)
\psline(-.6,0.4)(-0.6,-0.2)
\psline(0,-0.2)(0,0.4)
\psline(-0.6,-0.2)(0,-0.2)
&$483840$
\rule[-10pt]{0pt}{24pt}
\\

\hline
\psline[linearc=0.5]%
       (0.6,-0.2)(0.52,0.1)(0.6,0.4)
\psline[linearc=0.5]%
       (0.6,-0.2)(0.68,0.1)(0.6,0.4)
\psline(0,-0.2)(0,0.4)
\psline(0,0.4)(0.6,0.4)
\psline(-.6,0.4)(0,0.4)
\psline(-.6,0.4)(-0.6,-0.2)
\psline(-0.6,-0.2)(0,-0.2)
\psline[linearc=0.6]%
       (0.6,-0.2)(0.3,-0.12)(0,-0.2)
\psline[linearc=0.6]%
       (0.6,-0.2)(0.3,-0.28)(0,-0.2)
&$725760$
\rule[-9pt]{0pt}{23pt}
\\

\hline
\psline(0,-0.2)(0,0.4)
\psline(0.6,-0.2)(0.6,0.4)
\psline(-.6,0.4)(0,0.4)
\psline(-.6,0.4)(-0.6,-0.2)
\psline(-0.6,-0.2)(0,-0.2)
\psline[linearc=0.6]%
       (0.6,-0.2)(0.3,-0.12)(0,-0.2)
\psline[linearc=0.6]%
       (0.6,-0.2)(0.3,-0.28)(0,-0.2)
\psline[linearc=0.6]%
       (0.6,0.4)(0.3,0.32)(0,0.4)
\psline[linearc=0.6]%
       (0.6,0.4)(0.3,0.48)(0,0.4)
&$362880$
\rule[-9pt]{0pt}{24pt}
\\

\hline
\psline(0,-0.2)(0,0.4)
\psline(0.6,-0.2)(0.6,0.4)
\psline(-.6,0.4)(0,0.4)
\psline(-.6,0.4)(-0.6,-0.2)
\psline(0.6,0.4)(0,0.4)
\psline[linearc=0.6]%
       (0.6,-0.2)(0.3,-0.12)(0,-0.2)
\psline[linearc=0.6]%
       (0.6,-0.2)(0.3,-0.28)(0,-0.2)
\psline[linearc=0.6]%
       (-0.6,-0.2)(-0.3,-0.12)(0,-0.2)
\psline[linearc=0.6]%
       (-0.6,-0.2)(-0.3,-0.28)(0,-0.2)
&$362880$
\rule[-9pt]{0pt}{23pt}
\\

\hline
\psline(0,-0.2)(0,0.4)
\psline(0.6,-0.2)(0.6,0.4)
\psline(-.6,0.4)(0,0.4)
\psline(-.6,0.4)(-0.6,-0.2)
\psline(0.6,-0.2)(0,-0.2)
\psline[linearc=0.6]%
       (-0.6,-0.2)(-0.3,-0.12)(0,-0.2)
\psline[linearc=0.6]%
       (-0.6,-0.2)(-0.3,-0.28)(0,-0.2)
\psline[linearc=0.6]%
       (0.6,0.4)(0.3,0.32)(0,0.4)
\psline[linearc=0.6]%
       (0.6,0.4)(0.3,0.48)(0,0.4)
&$362880$
\rule[-9pt]{0pt}{24pt}
\\

\hline
\psline[linearc=0.5]%
       (0.6,-0.2)(0.52,0.1)(0.6,0.4)
\psline[linearc=0.5]%
       (0.6,-0.2)(0.68,0.1)(0.6,0.4)
\psline[linearc=0.5]%
       (-0.6,-0.2)(-0.52,0.1)(-0.6,0.4)
\psline[linearc=0.5]%
       (-0.6,-0.2)(-0.68,0.1)(-0.6,0.4)
\psline(0,-0.2)(0.6,-0.2)
\psline(0,0.4)(0.6,0.4)
\psline(-.6,0.4)(0,0.4)
\psline(0,0.4)(0,-0.2)
\psline(-0.6,-0.2)(0,-0.2)
&$181440$
\rule[-9pt]{0pt}{23pt}
\\
\hline

\end{tabular}

\begin{tabular}{|cc|}

\hline
\rule{18pt}{0pt}
\psline[linearc=0.6]%
       (0.6,0.4)(0.3,0.32)(0,0.4)
\psline[linearc=0.6]%
       (0.6,0.4)(0.3,0.48)(0,0.4)
\psline[linearc=0.5]%
       (-0.6,-0.2)(-0.52,0.1)(-0.6,0.4)
\psline[linearc=0.5]%
       (-0.6,-0.2)(-0.68,0.1)(-0.6,0.4)
\psline(0,-0.2)(0.6,-0.2)
\psline(-.6,0.4)(0,0.4)
\psline(0,0.4)(0,-0.2)
\psline(0.6,0.4)(0.6,-0.2)
\psline(-0.6,-0.2)(0,-0.2)
\rule[-9pt]{0pt}{24pt}
\rule{14pt}{0pt}
&$725760$
\\

\hline
\psline(0,-0.2)(0.6,-0.2)
\psline(-.6,0.4)(0,0.4)
\psline(0,0.4)(0,-0.2)
\psline(0.6,0.4)(0.6,-0.2)
\psline(-0.6,-0.2)(0,-0.2)
\psline(-0.6,0.4)(-0.6,-0.2)
\psline(0.6,.4)(0,0.4)
\psline[linearc=0.5]%
       (-0.6,-0.2)(-0.45,0.1)(-0.6,0.4)
\psline[linearc=0.5]%
       (-0.6,-0.2)(-0.75,0.1)(-0.6,0.4)
&$241920$
\rule[-9pt]{0pt}{23pt}
\\

\hline
\psline(0.6,0.4)(0,0.4)
\psline[linearc=0.6]%
       (-0.6,-0.2)(-0.65,0.1)(-0.6,0.4)
\psline[linearc=0.6]%
       (-0.6,-0.2)(-0.55,0.1)(-0.6,0.4)
\psline[linearc=0.4]%
       (-0.6,-0.2)(-0.77,0.1)(-0.6,0.4)
\psline[linearc=0.4]%
       (-0.6,-0.2)(-0.43,0.1)(-0.6,0.4)
\psline(0,-0.2)(0.6,-0.2)
\psline(-.6,0.4)(0,0.4)
\psline(0.6,0.4)(0.6,-0.2)
\psline(-0.6,-0.2)(0,-0.2)
&$181440$
\rule[-9pt]{0pt}{23pt}
\\

\hline
\psline(0.6,0.4)(0,0.4)
\psline(0,-0.2)(0.6,-0.2)
\psline(-.6,0.4)(-0.6,-.2)
\psline(0.6,0.4)(0.6,-0.2)
\psline(-0.6,0.4)(0,0.4)
\psline[linearc=0.5]%
       (-0.6,-0.2)(-0.45,0.1)(-0.6,0.4)
\psline[linearc=0.5]%
       (-0.6,-0.2)(-0.75,0.1)(-0.6,0.4)
\psline[linearc=0.6]%
       (-0.6,-0.2)(-0.3,-0.12)(0,-0.2)
\psline[linearc=0.6]%
       (-0.6,-0.2)(-0.3,-0.28)(0,-0.2)
&$725760$
\rule[-9pt]{0pt}{23pt}
\\

\hline
\psline(-0.6,0.4)(0,0.4)
\psline(0,.4)(0.6,.4)
\psline(-.6,0.4)(-0.6,-.2)
\psline(0.6,0.4)(0.6,-0.2)
\psline(-0.6,-0.2)(0,-0.2)
\psline[linearc=0.5]%
       (-0.6,-0.2)(-0.45,0.1)(-0.6,0.4)
\psline[linearc=0.5]%
       (-0.6,-0.2)(-0.75,0.1)(-0.6,0.4)
\psline[linearc=0.6]%
       (0.6,-0.2)(0.3,-0.28)(0,-0.2)
\psline[linearc=0.6]%
       (0.6,-0.2)(0.3,-0.12)(0,-0.2)
&$725760$
\rule[-9pt]{0pt}{23pt}
\\

\hline
\psline(-0.6,0.4)(0,0.4)
\psline(0,0.4)(0.6,0.4)
\psline(0,-0.2)(0.6,-0.2)
\psline(-.6,0.4)(-0.6,-.2)
\psline(-0.6,-0.2)(0,-0.2)
\psline[linearc=0.5]%
       (-0.6,-0.2)(-0.45,0.1)(-0.6,0.4)
\psline[linearc=0.5]%
       (-0.6,-0.2)(-0.75,0.1)(-0.6,0.4)
\psline[linearc=0.5]%
       (0.6,-0.2)(0.52,0.1)(0.6,0.4)
\psline[linearc=0.5]%
       (0.6,-0.2)(0.68,0.1)(0.6,0.4)
&$362880$
\rule[-9pt]{0pt}{23pt}
\\

\hline
\psline(-0.6,0.4)(0,0.4)
\psline(0,0.4)(0.6,0.4)
\psline(-.6,0.4)(-0.6,-.2)
\psline[linearc=0.5]%
       (0.6,-0.2)(0.52,0.1)(0.6,0.4)
\psline[linearc=0.5]%
       (0.6,-0.2)(0.68,0.1)(0.6,0.4)
\psline[linearc=0.6]%
       (0.6,-0.2)(0.3,-0.28)(0,-0.2)
\psline[linearc=0.6]%
       (0.6,-0.2)(0.3,-0.12)(0,-0.2)
\psline[linearc=0.6]%
       (-0.6,-0.2)(-0.3,-0.28)(0,-0.2)
\psline[linearc=0.6]%
       (-0.6,-0.2)(-0.3,-0.12)(0,-0.2)
&$544320$
\rule[-9pt]{0pt}{23pt}
\\

\hline
\psline(0.6,0.4)(0,0.4)
\psline(-.6,0.4)(-0.6,-.2)
\psline[linearc=0.5]%
       (0.6,-0.2)(0.52,0.1)(0.6,0.4)
\psline[linearc=0.5]%
       (0.6,-0.2)(0.68,0.1)(0.6,0.4)
\psline[linearc=0.6]%
       (0.6,-0.2)(0.3,-0.28)(0,-0.2)
\psline[linearc=0.6]%
       (0.6,-0.2)(0.3,-0.12)(0,-0.2)
\psline(-0.6,-0.2)(0,-0.2)
\psline[linearc=0.6]%
       (-0.6,0.4)(-0.3,0.32)(0,0.4)
\psline[linearc=0.6]%
       (-0.6,0.4)(-0.3,0.48)(0,0.4)
&$1088640$
\rule[-9pt]{0pt}{24pt}
\\

\hline
\psline[linearc=0.6]%
       (-0.6,0.4)(-0.3,0.32)(0,0.4)
\psline[linearc=0.6]%
       (-0.6,0.4)(-0.3,0.48)(0,0.4)
\psline(0.6,0.4)(0,0.4)
\psline(-.6,0.4)(-0.6,-.2)
\psline[linearc=0.5]%
       (0.6,-0.2)(0.52,0.1)(0.6,0.4)
\psline[linearc=0.5]%
       (0.6,-0.2)(0.68,0.1)(0.6,0.4)
\psline(0.6,-0.2)(0,-0.2)
\psline[linearc=0.6]%
       (-0.6,-0.2)(-0.3,-0.28)(0,-0.2)
\psline[linearc=0.6]%
       (-0.6,-0.2)(-0.3,-0.12)(0,-0.2)
&$181440$
\rule[-9pt]{0pt}{24pt}
\\

\hline
\psline(-.6,0.7)(0,0.7)
\psline(0.6,0.7)(0.6,0.1)
\psline(-0.6,0.7)(-0.6,0.1)
\psline(0.6,.7)(0,0.7)
\psline(-0.6,-0.5)(0,-0.5)
\psline(-0.6,-0.5)(-0.6,0.1)
\psline(0,-0.5)(0,0.1)
\psline[linearc=0.5]%
       (0,.1)(0.3,0.02)(0.6,.1)
\psline[linearc=0.5]%
       (0,.1)(0.3,0.18)(0.6,.1)
&$5806080$
\rule[-17pt]{0pt}{40pt}
\\

\psline(-.3,-0.8)(-0.3,1.0)
\psline(.3,-0.8)(0.3,0.4)
\psline(-.3,-0.8)(.3,-0.8)
\psline(-.3,1)(.3,1)

\psline[linearc=0.5]%
       (0.3,0.4)(0.22,0.7)(0.3,1)
\psline[linearc=0.5]%
       (0.3,0.4)(0.38,0.7)(0.3,1)
&$2903040$
\rule[-27pt]{0pt}{55pt}
\\

\psline(-.6,0.7)(0,0.7)
\psline(0.6,-0.5)(0,-0.5)
\psline(-0.6,0.7)(-0.6,0.1)
\psline(0.6,.7)(0,0.7)
\psline(-0.6,-0.5)(0,-0.5)
\psline(-0.6,-0.5)(-0.6,0.1)
\psline(0.6,-0.5)(0.6,0.1)
\psline[linearc=0.5]%
       (0.6,0.1)(0.52,0.4)(0.6,0.7)
\psline[linearc=0.5]%
       (0.6,0.1)(0.68,0.4)(0.6,0.7)
&$1451520$
\rule[-18pt]{0pt}{37pt}
\\

\hline
\psline(0,0.1)(0.6,0.1)
\psline(-.6,0.7)(0,0.7)
\psline(0,0.7)(0,0.1)
\psline(0.6,0.7)(0.6,0.1)
\psline(-0.6,0.7)(-0.6,0.1)
\psline(0.6,.7)(0,0.7)
\psline(-0.6,-0.5)(0,-0.5)
\psline(-0.6,-0.5)(-0.6,0.1)
\psline(0,-0.5)(0,0.1)
&$2903040$
\rule[-18pt]{0pt}{41pt}
\\

\psline(-.3,-0.8)(-0.3,1.0)
\psline(.3,-0.8)(0.3,1.0)
\psline(-.3,-0.8)(.3,-0.8)
\psline(-.3,1)(.3,1)
\psline(-.3,.4)(.3,0.4)
&$1451520$
\rule[-27pt]{0pt}{55pt}
\\

\hline
\end{tabular}
&

\end{tabular} \end{ruledtabular} \caption{List of all diagrams
contributing to the high temperature series expansion up to order
$\beta^8$.  All vertices with an odd number of lines correspond to a
$\sigma_i$ in the effective Ising model, to a $m$ in the mean-field
approximation.  $\mathcal{N}_g$ is the number  of ways in  the graph $g$
can be located on the lattice. } \label{tab:graphs} \end{table}
\subsection{Mean-field approximation}
The  expansion of Eq.~\ref{eq:Eeff} up   to $\beta^8$ leads  to a huge
number of terms and it is necessary   to proceed systematically in order
to obtain all the diagrams.  In mean-field, each chirality  is replaced
by its  mean value $\langle \sigma_i\rangle=m$, which  simplifies  the
diagrammatic   expansion of  the   effective Hamiltonian. We   have
written a (Maple)   symbolic code that i) generates  all  possible
diagrams  on   the square lattice, ii) assigns a weight $m$ to multiply
connected vertices with an  odd number of bonds iii) computes the
integrals over the different  vectors  exactly    and iv) computes  the
number of ways $\mathcal{N}_g$   the  graph $g$   can  be  located on
the lattice.  $\mathcal{N}_g$ corresponds to the product  of the number
of bond  ordering  by     the number  of transformations (rotations,
reflections, deformations on  the lattice) changing the representation
of the graph but not the graph itself. It is also worth noting that
graphs free of articulation vertex  have   zero cumulants.  In
addition,  we discard graphs  that   give irrelevant constant
contributions, \textit{i.e.}   graphs where   the multiplicity of each
vertex   is even. With  these prescriptions, only $55$  graphs
contribute at order $\beta^8$. They are listed in Tab.\ref{tab:graphs}.

% At  order $\beta^8$, the effective  multiple-spin Ising model involves
% up to six spin interactions (see Table \ref{tab:graphs})

\begin{figure}
\includegraphics[width=7cm]{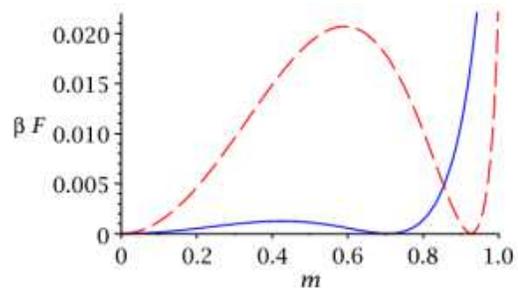}
\caption{(Color online) Mean-field free energy for model~(\ref{eq:p}) at
the transition temperature, for $p=1$ (Ising-$\mathbb{R}P^3$ model,
solid line) and $p=2$ (dashed line).  In both cases there are two stable
mean-field solutions and the transition is first order. However the
energy barrier is much smaller for $p=1$.}
\label{fig:Fchpmoy}
\end{figure}

As a result, we obtain the effective energy in the mean-field
approximation:
\begin{eqnarray}
E_{\textrm{eff}}(m)&=&
\left(-\frac{\beta^2}{3}-\frac{7\beta^4}{45}-\frac{\beta^5}{9}-\frac{391\beta^6}{4536}-\frac{\beta^7}{81}
\right. \nonumber\\ && \left.  +\frac{5173\beta^8}{145800}\right) m^2
\nonumber \\
&&+\left(-\frac{\beta^5}{18}-\frac{7\beta^7}{162}-\frac{2\beta^8}{27}\right)m^4\nonumber
\\ &&+\left(-\frac{\beta^8}{81}\right)m^6.
\label{eq:EMF}
\end{eqnarray}
Hence the mean-field free energy is  given by $F(m)=E(m)-TS(m)$ where
the  entropy
$S(m)=-\frac{1+m}{2}\ln\left(\frac{1+m}{2}\right)-\frac{1-m}{2}\ln\left(\frac{1-m}{2}\right)$.
$F(m)$ is minimized  with respect to the magnetization $m$: a  phase
transition occurs  at  $T=1.07$ between an  ordered phase $m\ne 0$ and a
paramagnetic phase $m=0$. Further, the transition is of first order
albeit with a very small free  energy  barrier (see
Fig.~\ref{fig:Fchpmoy}).

This computation is repeated for the Ising-$\mathbb{R}P_3$         model
with   ``non-linear'' interaction, (Eq.~\ref{eq:p} with $p=2$). We
obtain
\begin{eqnarray}
E_{\textrm{eff}}(m)&=&
\left(1-\frac{\beta^2}{3}-\frac{81\beta^4}{800}-\frac{13\beta^5}{200}-\frac{1281493\beta^6}{10080000}
\right.\nonumber\\ &&\left.
-\frac{4877\beta^7}{80000}-\frac{5746857169\beta^8}{82944000000} \right)
m^2 \nonumber \\ &&+\left(
-\frac{13\beta^5}{400}-\frac{1351\beta^7}{32000}-\frac{227\beta^8}{8000}
\right)m^4\nonumber \\ &&+\left( -\frac{227\beta^8}{48000}  \right)m^6.
\label{eq:EMFp2}
\end{eqnarray}
As can be seen in Figure~\ref{fig:Fchpmoy}, the first-order transition
in this model is much stronger than in the $p=1$ case, {\it i.e.}
$(\Delta \beta_c F)_{p=2} \gg (\Delta \beta_c F)_{p=1}$.
We also mention that similar conclusions can be made for
model~(\ref{eq:a}) with $a=1.75$.

Once again, these results support our claim that  the
Ising-$\mathbb{R}P^3$ model is close to a point in parameter space
where the transition changes from  first to second order, in qualitative
agreement with the simulation results.
\section{Conclusion}\label{sec:conclusion}

We have proposed a  minimal Ising-$\mathbb{R}P^3$ model that  captures
most of  the low energy  features of frustrated Heisenberg models with
an $O(3)-$manifold of  ground states.  The discrete  symmetry-breaking
associated  with the existence of  two  connected components of $O(3)$
leads to  an  Ising-type   continuous transition with   unconventional
features at small sizes.  The complexity  of this transition is due to
the strong coupling of the Ising chirality fluctuations to short range
continuous spin  fluctuations and to  the topological defects of these
textures.

We provided a consistent  picture of  these $\Z_2$ defects in presence of
chiral fluctuations. Namely, we showed that chiral domain walls can also
carry a $\Z_2$ topological charge, albeit delocalized over the
interface.  Inspection of configurations revealed that isolated vortex
cores  are almost  always  located nearby charged domain walls.  At the
transition temperature, most defects consist in vortex/charged-wall
pairs, while there are a few vortex/vortex pairs and almost no isolated
defect.  This mechanism is essential in understanding the appearance  of
point defects  at  a temperature much lower   than in the case where
chiral fluctuations are absent.

We also studied some    variants of the    model and showed  that  the
continuous  transition easily becomes  first-order,  which lead us  to
conjecture  the existence of  a nearby tricritical  point in parameter
space.   We clarified  the role   of  short-range fluctuations of  the
continuous variables  in this mechanism  by an analogy  with the large
$q$-state Potts model. In this analogy the high density of states with
orthogonal  4D-vectors translate into    the large number $q(q-1)$  of
states with  disordered bonds of  the Potts model.  Finally we studied
two derived versions of the original model:  i) a diluted-Ising model,
which   somehow  corresponds  to   a  coarse-grained version    of the
Ising-$\mathbb{R}P^3$  model  (the  continuous  variables are  locally
averaged and replaced  by discrete  variables),  and ii)  an effective
multispin Ising model, obtained by tracing out the vector spins, order
by order  in $\beta$  (up to  $\beta^8$).   These two   models predict
either a weakly first-order or a continuous transition, as well as the
existence of a tricritical point.

This study sheds light on the large variety  of behaviors reported for
chiral       phase     transitions       in       frustrated      spin
systems.~\cite{momoi97,cs04,domenge05,domenge08} In all  these models,
the   chirality  is an  emergent   variable  more or  less  coupled to
short-range  spin fluctuations: in the $J_1-J_3$  model  on the square
lattice the chiral variable is the pitch of an  helix, the coupling to
the short range spin fluctuations is probably small and the transition
appears to be  clearly   second-order and in the    Ising universality
class.~\cite{cs04}  In  the   cyclic 4-spin  exchange   model  on  the
triangular  lattice,  the  order parameter is    a tetrahedron and the
chiral variable is associated to the  triple product of three of these
four     spins: the  transition    is   probably   very  weakly  first
order.~\cite{momoi97,momoi99} In  the  $J_1-J_2$ model  on  the kagom\'e
lattice, the   order parameter at $T=0$   is  a cuboactedron,  and the
chiral variable is associated to the triple product  of three of these
twelve  spins.  The phase  transition evolves from  weakly to strongly
first     order    when     tuning    the    parameters    towards   a
ferromagnetic-antiferromagnetic phase  boundary at $T=0$: this  can be
understood in  the  light of the present  work.  Tuning the parameters
towards the    ferromagnetic phase frustrates   the 12-sublattice Néel
order and  favors short-range  disorder    and vortex formation.   The
associated increase in  the entropy  unbalance drives the   transition
from Ising  to   strongly first order,   much in  the same way   as in
Section~\ref{sec:potts}.

On the technical side, this proximity of a     tricritical point in
parameter   space evidences   why simulations and  experiences  must
be  lead with great caution, a conclusion equally supported by recent
work from the quite different standpoint of the non perturbative
renormalisation group approach.~\cite{dmt04}

\section{Acknowledgments}
We thank D.  Huse for  suggesting this  work and B.  Bernu, E. Brunet,
R. Mosseri and V. Dotsenko for helpful discussions.
\appendix

\section{Monte-Carlo Algorithm}\label{sec:monte-carlo-algor}
In   this          section     we    detail        the     Wang-Landau
algorithm~\cite{Wang2001,Schulz2003}    used        to        simulate
model~(\ref{eq:energy}). This method  consists in building the density
of states   $g(E)$   progressively,   using  successive    Monte Carlo
iterations. Elementary moves   consist in rotations  of  vectors $\vec
v_i$ as well as flips of chiralities $\sigma_i$. In the case where the
chiralities are frozen (Section~\ref{sec:fixed-chir-limit}) only   the
rotation movements are  performed.  For  completeness we mention  that
the four-vectors $\vec  v_i$ are sampled  uniformly  on $S^3$ using  3
random numbers ($r$,  $\eta$  and  $\nu$), independent and   uniformly
distributed in $[0,1]$, according to
\begin{equation}
v_1=\begin{cases} \sqrt{r}\cos(2\pi \eta)&\\ \sqrt{r}\sin(2\pi\eta)\\
  \sqrt{1-r}\cos(2\pi \nu)&\\ \sqrt{1-r}\sin(2\pi\nu)
\end{cases}
\end{equation}
Starting  from an  initial guess $g(E)$, the acceptance of a trial
flip/rotation is decided by a Metropolis rule
\begin{equation}
\Pi(o\rightarrow n)={\rm Min}\left(1,\frac{g(E_o)}{g(E_n)}\right)
\end{equation}
where   the subscripts  $o$  and  $n$ corresponds    the old and new
configurations,  respectively.

Every time a configuration with energy $E$ is  visited, the density of
states $g(E)$ is multiplied by a factor $f>1$, $g(E)\leftarrow g(E)f$.
To ensure that all configurations are well sampled, a histogram $H(E)$
accumulates all  visited states. The first part  of the run is stopped
when $H(E)>10^4$.  In a second part, the histogram $H(E)$ is reset and
the run is continued, but the modification factor $f$ is now decreased
to $f_1<f$. In the  original paper by Wang,~\cite{Wang2001} $f_1$  was
taken  as $\sqrt{f}$, but this choice  is not necessarily the best for
continuous  models.~\cite{Poulain2006}   In this  case the convergence
properties  were found   to   be   quite  satisfying   upon   choosing
$f_1=f^{0.7}$.   The random walk  is continued  until the histogram of
visits $H(E)$  has become ``flat''.  Once  again, $H(E)$ is  reset and
the modification factor is decreased   to $f_2<f_1$, etc...   Accurate
density  of  states  $g_n(E)$ are  generally  obtained   at the $n$-th
iteration, where $n$  is  such  that   $f_n$ is  almost 1   (typically
$f_n-1\lesssim10^{-8}$).

Since this model has continuous variables, its energy spectrum is also
continuous, and the choice of the energy bin requires special care: if
the bin is too large,  important details of the  spectrum may be lost,
whereas if it is too small, a lot of computer time is wasted to ensure
the convergence of the method.   In the temperature range of interest,
a size of order $\simeq  0.1$ is a  good compromise  for the model  of
Eq.~\ref{eq:energy_vv2}.  In addition, the  energy range is limited to
the  region relevant at the  transition, and  yields all thermodynamic
quantities accurately at these temperatures.

Once the density of states $g(E)$ is obtained, all moments of the energy
distribution can be computed in a straightforward way as
\begin{equation}
\langle E^n\rangle=\frac{\int g(E)E^n\exp(-\beta E)}{\int
g(E)\exp(-\beta E)}
\end{equation}
from which the specific heat per site is readily obtained:
\begin{equation}
C_v=\frac{1}{N}\left(\langle E^2\rangle-(\langle E\rangle)^2\right).
\end{equation}
For the thermodynamic quantities that are not directly   related to
the moments  of the energy distribution, such as the chirality, the
vorticity, and their associated susceptibilities, or the Binder cumulants,
we proceed as follows: an additional simulation is performed where
$g(E)$ is no longer modified (a  ``perfect''  random walk in  energy
space   if the density of states $g(E)$ is  very accurate). In this last
run additional histograms are stored: chirality   histograms
$\sigma(E)$, $\sigma^2(E)$, $\sigma^4(E)$, and vorticity histogram
$V(E)$, $V^2(E)$, $V^4(E)$.

Chirality   (or vorticity) moments  are     then  obtained   from the
simple 1D integration
\begin{equation}
\langle M^n\rangle=\frac{\int g(E)\frac{M^n(E)}{H(E)}\exp(-\beta
E)}{\int g(E)\exp(-\beta E)}
\end{equation}
from which, say, the Binder cumulant, is readily obtained as
\begin{equation}
U=1-\frac{\langle M^4 \rangle}{3(\langle M^2 \rangle )^2}
\end{equation}
where $M=\sigma$ or $V$.  Contrary to the method of the Joint Density of
States,~\cite{Zhou2006} our method does not require the construction of
the  two dimensional histogram $g(E,\sigma)$. Hence it is not limited to
modest lattice sizes (the above quantities are computed for $L$ up to $L=80$).

\end{document}